\documentclass[a4paper,onecolumn,longbibliography,accepted=2023-03-15]{quantumarticle}
\pdfoutput=1

\usepackage[margin=1in]{geometry} 
\usepackage{hyperref}
\usepackage[numbers, sort&compress]{natbib}
\usepackage[utf8]{inputenc}
\usepackage{braket}  
\usepackage{amsmath} 
\usepackage{amssymb} 
\usepackage{amsthm}  
\usepackage{mathtools} 
\usepackage{xcolor, soul} 
\usepackage{verbatim} 

\newcommand{\pth}{\psi_\theta} 
\newcommand{\ketbra}[1]{\ket{#1}\bra{#1}} 
 
\newcommand{\underi}[1]{{i}_{\underline{#1}}}

\newtheorem{theorem}{Theorem}[section]
\newtheorem{proposition}[theorem]{Proposition}
\newtheorem{lemma}[theorem]{Lemma}

\newtheorem{remark}{Remark}

\begin{document}
\title{Adaptive measurement filter: efficient strategy for optimal estimation of quantum Markov chains}
\author{Alfred Godley and M\u{a}d\u{a}lin Gu\c{t}\u{a}}


\affiliation{School of Mathematical Sciences, University of Nottingham, University Park, NG8 2SD Nottingham, United Kingdom}

\maketitle


\begin{abstract}
Continuous-time measurements are instrumental for a multitude of tasks in quantum engineering and quantum control, including the estimation of dynamical parameters of open quantum systems monitored through the environment. However, such measurements do not extract the maximum amount of information available in the output state, so finding alternative optimal measurement strategies is a major open problem. 

In this paper we solve this problem in the setting of 
discrete-time input-output quantum Markov chains. We present an efficient algorithm for optimal estimation of one-dimensional dynamical parameters which consists of an iterative procedure for updating a `measurement filter' operator and determining successive measurement bases for the output units. A key ingredient of the scheme is the use of a coherent quantum absorber as a way to physically `post-process' the output after the interaction with the system. The absorber is designed such that the joint system plus absorber stationary state is pure at parameter value provided by a preliminary estimator. The scheme offers an exciting prospect for optimal continuous-time adaptive measurements, but more work is needed to find realistic practical implementations.

\end{abstract}

\section{Introduction}
\label{sec:Introduction}

The quantum input-output (I-O) formalism is an effective framework for describing the evolution, monitoring and control of Markovian quantum open systems \citep{GZ04,WM10,CKS17}. In this setting, the interaction with the environment is modelled by coupling the system of interest with a quantum transmission line (channel) represented by a Gaussian bosonic field. 
The output field carries information about the open system's dynamics which can be accessed by performing continuous-time measurements, and the corresponding conditional system evolution is described in terms of stochastic Schr\"{o}dinger or filtering equations \citep{Belavkin94,BvHJ07,Carmichael,DCM,WM93}.


While these theories are key to quantum engineering applications, they rely on the precise knowledge of the system's dynamical parameters (e.g. Hamiltonian of field coupling), which are often uncertain, or completely unknown, and therefore  need to be estimated from measurement data. The I-O formalism is ideally suited for this statistical inference task, and more generally for implementing online system identification methods \citep{Ljung99}. Unlike direct measurement techniques which require repeated system re-preparations and fast control operations \citep{Huelga97,Escher11,DemkowiczGuta12,Benatti14,Smirne16,Demko17,Zhou18}, the parameters can be estimated continuously from the output measurement trajectory, even if the system is not directly accessible or it is involved in an information processing task. The first investigation in parameter estimation for continuously-observed quantum systems considered the estimation of the Rabi frequency of an atom in a cavity mode, while a photon counting measurement is performed on the cavity output \citep{Mab96}. 
Subsequent works have addressed a variety of related problems including the dependence on measurement choice \citep{GW01}, adaptive estimation \citep{Berry02,Wiseman04} filtering with uncertain parameters \citep{Ralph11} particle filters for estimation \citep{CG09,Six15} achieving Heisenberg scaling \citep{GutaMacieszczakGarrahanLesanovsky,ARPG17,Genoni18}, sensing with error correction \citep{Plenio16}, Bayesian estimation \citep{GM13,Negretti13,KM16,Ralph17,Zhang19}, quantum smoothing \citep{Tsang09,T09,Tsang10,Guevara15}, waveform estimation \citep{TWC11,Tsang2012} estimation of linear systems \citep{GY16,Genoni17,Levitt17,LGN18}, classical and quantum Fisher informations of the output channel \citep{Guta2011,GM13,GutaCatanaBouten,KM14,Molmer2014,Genoni17,GutaKuikas,GK17}.

An upshot of these studies is that standard measurements such as counting, homodyne or heterodyne generally do no achieve the ultimate limit given by the quantum Cramér–Rao bound \citep{Helstrom1976,Holevo,BraunsteinCaves}, while the optimal measurement prescribed by the symmetric logarithmic derivative requires collective operations on the output state. In this work we make a first step towards addressing the key issue of devising \emph{realistic and statistically effective} measurement strategies within the the framework of the I-O theory. By realistic we mean procedures which involve sequential continuous-time measurements (as opposed to general non-separable measurements on the output state), possibly combined with more advanced but theoretically well understood operations such as series connections and feedback \citep{GJ09}.

\begin{figure}[h]
        \centering
        \includegraphics[width=7cm]{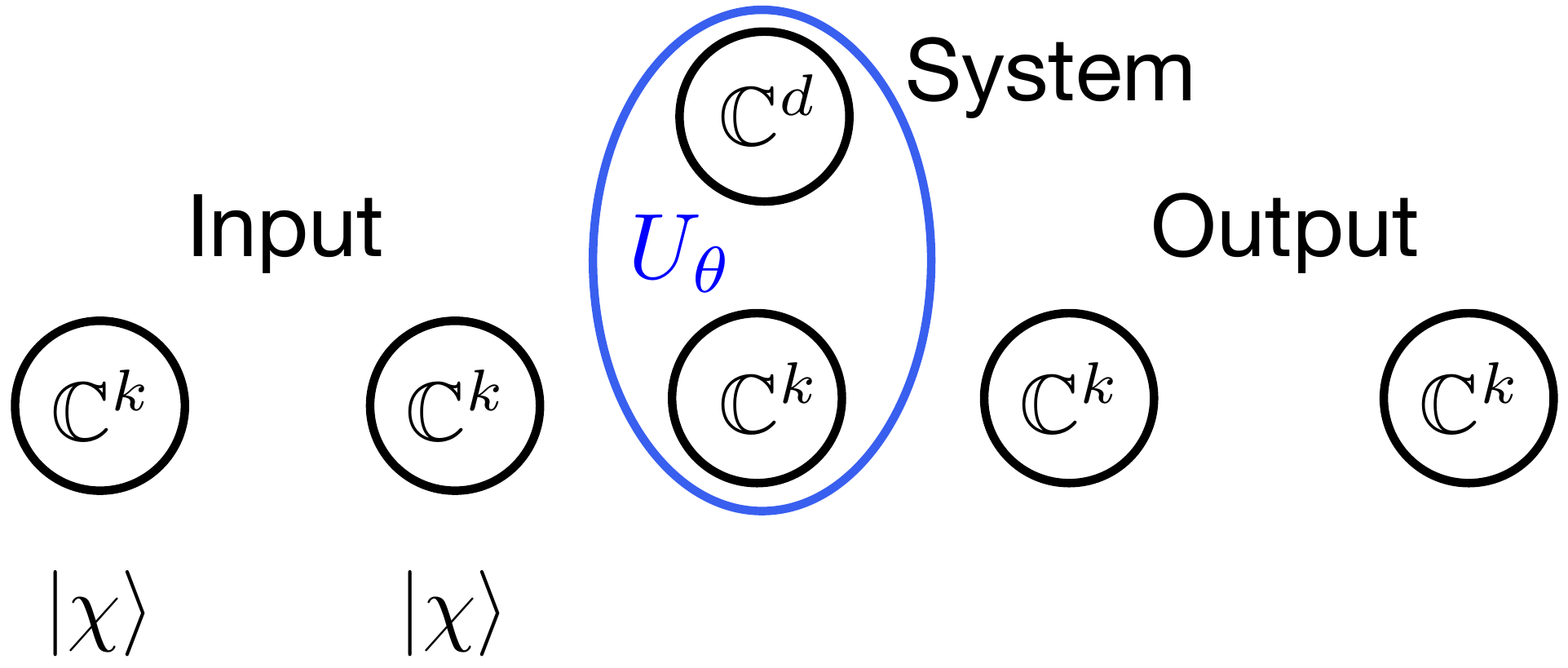}
        \caption{Quantum input-output discrete-time dynamics with $\theta$ dependent unitary interaction $U_\theta$}
        \label{Fig:i-o}
    \end{figure}

For conceptual clarity we focus primarily on discrete-time dynamics, but we will indicate how the techniques may be extended to continuous-time. In the discrete-time setting, the I-O dynamics consists of a $d$-dimensional system of interest interacting sequentially with a chain of $k$-dimensional `noise' input units, which are identically and independently prepared in a state $|\chi\rangle$, cf. Figure \ref{Fig:i-o}. 
We assume that the interaction unitary $U_\theta$ acting on $\mathbb{C}^d\otimes \mathbb{C}^k $  depends on a parameter  $\theta\in \mathbb{R}$, which we would like to estimate by measuring the output after $n$ evolution steps. In principle, this can be done by applying the adaptive, separable measurement scheme 
developed in \citep{Zhou20} to the joint pure system-output state; indeed this has been shown to attain the quantum Cram\'{e}r-Rao bound. However, while theoretically applicable, the algorithm involves manipulating multi-partite operators, making it unsuitable for processing output states with a large number $n$ of noise units. In addition, it is not clear how the algorithm can be applied to continuous-time dynamics.

Our main contribution is to eliminate these drawbacks by devising a scheme which exploits the intrinsic Markovian structure of the problem. 
Concretely, we propose an algorithm which finds optimal measurement bases for each of the output units by only performing computations on the space of a doubled-up system and a noise unit, i.e. $\mathbb{C}^d\otimes \mathbb{C}^d\otimes \mathbb{C}^k$. Our algorithm has a similar structure to that of the quantum state filter describing the system's conditional evolution, and can be run in real-time without having to specify the time length $n$ in advance. 

While our general algorithm requires measurements on both the output and system in order to achieve finite sample optimality, in Proposition \ref{prop:CFI-QFI} we prove that by measuring only the output we incur a loss of Fisher information which is bounded by a constant, independent of the time $n$. Since the quantum Fisher information scales linearly in time, this implies that output measurement is optimal in the leading contribution to the QFI. 

We now describe our scheme in more detail. In the first stage of the protocol we use a small proportion of the output units (of sample size $\tilde{n}\approx n^{1-\epsilon}$ with small $\epsilon$) in order to compute a preliminary `rough estimator' $\theta_0$ of the true parameter $\theta$, by performing a standard sequential measurement. This step is necessary in \emph{any} quantum estimation problem in which the optimal measurement depends on the unknown parameter \citep{GillMassar}. In particular, this means that strictly speaking one can only attain optimality in the limit of large sample sizes, as $\theta-\theta_0$ decays as $n^{-(1-\epsilon)/2}$ thanks to the preliminary estimation stage.
\begin{figure}[h!]
 \centering
        \includegraphics[width=12cm]{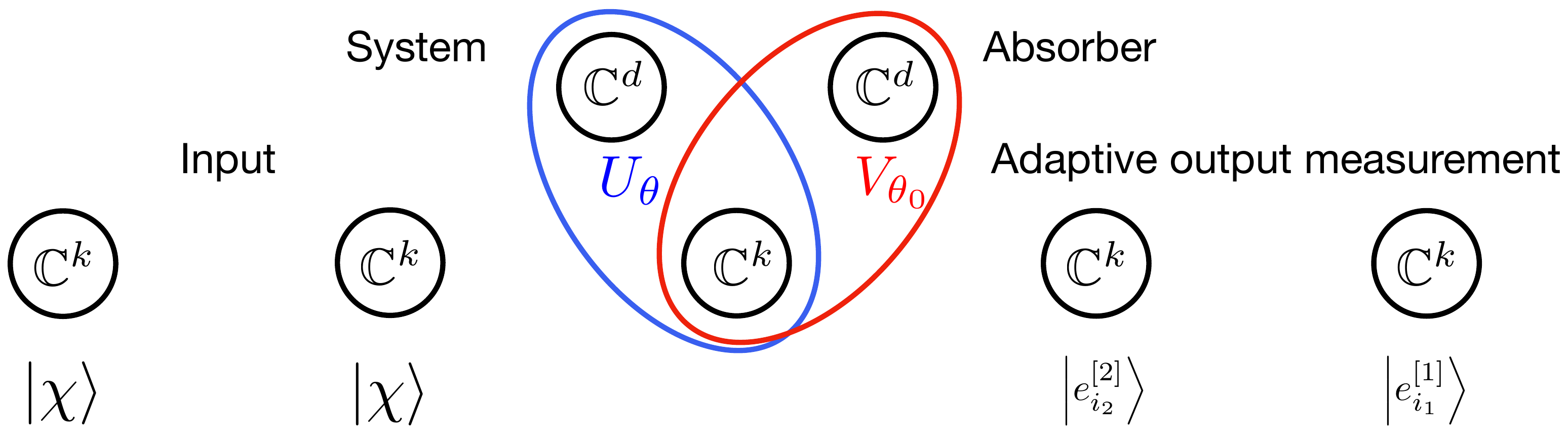}
        \caption{Adaptive measurement filter:  the output units undergo post-processing with a coherent quantum absorber, followed by applying an adaptive measurement computed with the algorithm}
        \label{Fig:adapt-meas}
\end{figure}
In the second (main) stage of the protocol we use $\theta_0$ to design a sequential measurement which achieves the output QFI at $\theta =\theta_0$. Since the first stage insures that $\theta-\theta_0$ decays and the QFI is continuous with respect to $\theta$, we find that the overall scheme is asymptotically optimal at any parameter value $\theta$. 

The second stage is illustrated in Figure \ref{Fig:adapt-meas}: each output unit undergoes a physical transformation (which we call `quantum post-processing') followed by an adaptive projective measurement whose basis is computed according to the `measurement filter' algorithm described below. More specifically, after interacting with the system, the post-processing consists in applying a unitary $V_{\theta_0}$ to the output noise unit together with an additional ancilla of the same size $d$ as the system. The system and ancilla can be regarded as a single open system (denotes `s+a') of dimension $D= d^2$ coupled to the noise units via the unitary $W_\theta = V_{\theta_0}
U_\theta$. 
The unitary $V_{\theta_0}$ is chosen such that s+a has a \emph{pure stationary state} $|\psi\rangle\in \mathbb{C}^D$ at $\theta=\theta_0$, and the output state is identical to that of the input. This is a discrete-time analogue of the notion of \emph{coherent quantum absorber} introduced in \citep{Preskill}, and it insures that the `reference' output state at $\theta=\theta_0$ is the same as the product input state (the `vacuum'), while deviations from $\theta_0$ produce 'excitations' in the output. After the interaction with the ancilla (absorber), the noise unit is measured in a basis determined by a simple iterative algorithm 
detailed in section \ref{sec:adaptive.measurement}. 
The iterative step consists of using the current value of a certain `filter operator' on system+absorber to determine the next measurement basis, and then using the measurement outcome to update the filter operator. This simplification relies on the fact that the output is uncorrelated from system (and absorber), which is not the case in the original dynamical setup of Figure \ref{Fig:i-o}.

In section \ref{sec:Simulations} we describe the results of two numerical investigations testing our theoretical results. The first investigation focuses on a simplified model where the system plus absorber are represented by a two-dimensional system with a pure stationary state. While this sidesteps the preliminary estimation stage of the protocol, it allows us to specifically test the key features of the adaptive measurement algorithm with a reasonably large trajectory length and a high number of repetitions. For this model, we can explicitly compute the system-output QFI (cf. Lemma \ref{lemma:QFI-finite-time}), while the classical Fisher information of any output measurement strategy can be estimated by sampling techniques. The results confirm that the adaptive measurement attains the QFI when the system is measured at the end, while the output-only strategy is only worse by a constant independent of trajectory length. On the other hand, simple measurements (same fixed basis for each unit) perform strictly worse even when the measurement basis is optimised. While the improvement here is not dramatic, our preliminary investigations indicate that the gap increases significantly with the system dimension, depending on the chosen model. We further test the performance of the maximum likelihood estimator and find that its mean square error approaches the inverse of the classical Fisher information in the long-time limit, which agrees broadly with the Cram\'{e}r-Rao bound. The second numerical investigation implements the full two-stage adaptive measurement algorithm including the use of the coherent absorber.

Finally, we note that our scheme can be extended to continuous-time dynamics by using standard time-discretisation techniques \citep{Osborne,Horssen13}. Although we do not treat this in detail here, we comment on this extension at the end of the paper.


The paper is organised as follows. In section \ref{sec:Jiang} we briefly review the adaptive algorithm for optimal separable measurements developed in \citep{Zhou20}. Section \ref{sec:Markov} introduces the  Markov dynamics setting and reviews a key result on the asymptotic QFI of the output. Section \ref{sec:pure.stationary} explains how the use of `post-processing' by quantum absorber reduces the general estimation problem to one concerning a system with a pure stationary state. This is then used in section \ref{sec:adaptive.measurement}, which details the adaptive measurement procedure including the key `measurement filter' algorithm. In section \ref{sec:classical_Fisher} we show that the proposed adaptive output measurement achieves the optimal QFI rate even if the system is not measured. We also devise a scheme to estimate the classical Fisher information of the measurement process by sampling over trajectories. Section \ref{sec:Simulations} presents simulation results using an elaboration of an amplitude decay qubit model.

\section{Optimal separable measurements}
\label{sec:Jiang}

In this section we review a result by
Zhou, Zou and Jiang \citep{Zhou20} concerning optimal parameter estimation for multipartite pure states, using separable measurements (local measurements and classical communication). Their method will then be applied to the problem of estimating parameters of discrete time quantum input-output systems. By exploiting the 
Markovian nature of the dynamics, we will show that the algorithm can be recast in a simpler procedure akin to that of a quantum state filter. 

Consider a one parameter quantum statistical model $\{ \rho_\theta :\theta\in \mathbb{R}\}$ where $\rho_\theta$ is a state on a Hilbert space $\mathcal{H}$ which depends smoothly on the unknown parameter $\theta$. To estimate $\theta$ we perform a measurement on the state $\rho_\theta$ and compute an estimator $\hat{\theta}$ based on the measurement outcome. According to the quantum Cram\'{e}r-Rao bound (QCRB) \citep{Helstrom1976,Holevo,BraunsteinCaves}, the variance of any unbiased estimator $\hat{\theta}$ is lower bounded as
$$
{\rm Var}(\hat{\theta}) = \mathbb{E}\left[(\hat{\theta} -\theta)^2\right]\geq F_\theta^{-1}
$$
where $F_\theta$ is the quantum Fisher information (QFI) defined as $F(\theta) = {\rm Tr} (\rho_\theta \mathcal{L}_\theta^2)$, and $\mathcal{L}_\theta$ is the symmetric logarithmic derivative (SLD) satisfying $\frac{d}{d\theta} \rho_\theta = \mathcal{L}_\theta \circ \rho_\theta$. 
In general, for any given parameter value $\theta_0$, the QCRB is saturated\footnote{ This achievability argument can be made rigorous in an asymptotic setting where the experimenter has an ensemble of $n$ independent, identically prepared systems and employs an adaptive procedure for `locating' the parameter \citep{GillMassar,DemkowiczGuta}. } by measuring the SLD $\mathcal{L}_{\theta_0}$ and constructing a locally unbiased estimator 
$\hat{\theta} = \theta_0 + X/F(\theta_0)$ where $X$ is the measurement outcome.

While for full rank states the optimal measurement is essentially unique, for rank deficient states this is not the case and a necessary and sufficient condition for a measurement to saturate the QCRB has been derived in \citep{BraunsteinCaves}. This has practical relevance for multipartite systems where the measurement of the SLD may not be easy to implement. Motivated by this limitation, the  saturability condition has been further investigated in \citep{Zhou20}  where it is shown that the QCRB for pure states of multipartite systems is achievable using separate measurements constructed in an adaptive fashion which we now proceed to describe.

Consider the pure state model $\rho_\theta=\ketbra{\pth}$ with 
$|\psi_\theta\rangle\in \mathcal{H}$. We denote $|\dot{\psi}_\theta\rangle= \frac{d}{d\theta} |\psi_\theta\rangle$ and assume that 
$
\langle \psi_\theta| \dot{\psi}_\theta\rangle=0.
$
This can generally be arranged by choosing the (unphysical) phase of $|\psi_\theta\rangle$ to have an appropriate dependence on $\theta$. In particular, in this case we have $\ket{\pth^\perp}:=(1-\ketbra{\pth}) \ket{\dot{\pth}}= \ket{\dot{\pth}}$. Under this assumption the QFI is given by 
\begin{equation}\label{eq:QFI.pure}
F_\theta = 4 \|\dot\psi_\theta\|^2. 
\end{equation}
Further, we define the operator $M$ which will play a key role in the analysis
\begin{equation}
    M=\ket{\pth}\bra{\dot{\pth}} - \ket{\dot{\pth}}\bra{\pth}.
\end{equation}
The authors of \citep{Zhou20} note that if a projective rank-one measurement $\{E_i = |e_i\rangle \langle e_i|\}$
satisfies the conditions
\begin{equation}\label{eq:condition.Jiang}
\langle e_i |M |e_i\rangle =0, \qquad {\rm and}\qquad p_\theta(i)= |\langle e_i|\psi_\theta\rangle|^2 = 1/k, \qquad k= {\rm dim}(\mathcal{H})
\end{equation}
then it fulfils the general criteria of \citep{BraunsteinCaves} and therefore it achieves the QCRB. In fact, the second condition can be relaxed to $p(i)\neq 0$ for all $i$, but we will stick to the chosen expression for concreteness. The achievability can be understood as follows. The conditions $\langle e_i |M |e_i\rangle =0$ implies that $\langle e_i |\psi_\theta\rangle \langle \dot{\psi}_\theta |e_i\rangle$ is real, so that the phase of the basis vectors $|e_i\rangle$ can be chosen such that both 
$\langle e_i |\psi_\theta\rangle$ and $\langle e_i |\dot{\psi}_\theta\rangle$ are real for all $i$. Together with the condition $p_\theta(i)\neq 0$, this means that in the first order of approximation, the quantum model is described by vectors with real coefficients with respect to the measurement basis. In this case the classical and quantum informations coincide
\begin{eqnarray*}
I_\theta &=& \sum_i p_\theta(i)\left(\frac{d\log p_\theta(i)}{d\theta}\right)^2 =
4 \sum_i \frac{\left({\rm Re}\langle e_i |\psi_\theta\rangle \langle \dot{\psi}_\theta |e_i\rangle\right)^2 }{\langle e_i |\psi_\theta\rangle \langle \psi_\theta |e_i\rangle}\\
&=& 4 \sum_i \frac{\left|\langle e_i |\psi_\theta\rangle \langle \dot{\psi}_\theta |e_i\rangle\right|^2 }{\langle e_i |\psi_\theta\rangle \langle \psi_\theta |e_i\rangle} = 4\sum_i \left|\langle e_i |\dot{\psi}_\theta\rangle\right|^2 = 4\|\dot{\psi}_\theta||^2= F_\theta .
\end{eqnarray*}

We now assume that we deal with a multipartite system such that $\mathcal{H} = \mathcal{H}_1 \otimes \mathcal{H}_2 \otimes \dots\otimes \mathcal{H}_n$, with ${\rm dim}(\mathcal{H}_i) =k_i$, and follow \citep{Zhou20} to show that a separable measurement satisfying the above conditions can be constructed by using the algorithm outlined below. For any set of indices $A\subset \{1,\dots, n\}$ we denote its complement by $A^c$ and by $\rho_A = {\rm Tr}_{A^c} (\rho_\theta)$ the partial state of the sub-systems with indices in $A$. For $m>1$ we denote by $\underline{m}$ the set $\{1,\dots, m\}$.  Similarly, we denote $M_A =  {\rm Tr}_{A^c} (M)$ and for single sub-systems ($A= \{i\}$) we use the notation $\rho_i$ and  $M_i$.

We measure the sub-systems sequentially, such that each individual measurement basis depends on the outcomes of the previous measurements, as follows. In the first step, the measurement basis $\{|e_i^{[1]}\rangle\}_{i=1}^{k_1}$ 
of system $\mathcal{H}_1$ is chosen such that 
$$
\left\langle e_i^{[1]}\right| M_1\left| e_i^{[1]}\right\rangle =0  \qquad {\rm and} \qquad 
p_1(i) = \left\langle e_i^{[1]}\right|\rho_1\left| e_i^{[1]}\right\rangle = \frac{1}{k_1}.
$$
The existence of such a basis can be established by induction with respect to dimension, cf. proof of Lemma 1 in \citep{Zhou20}. The concrete construction in two dimensions is described in section \ref{sec:Simulations}.  


After this, the following procedure is applied sequentially to determine the measurement basis for system $j+1$ with $j=1,\dots,  n-1$: given the outcomes $i_{\underline{j}}:=\{i_1,\dots, i_j\}$ of the first $j$ measurements, we choose the basis $\left\{|e^{[j+1]}_i \rangle \right\}_{i=1}^{k_{j+1}}$ in $\mathcal{H}_{j+1}$ such that 
$$
\left\langle e_{i}^{[j+1]} \right| M_{j+1}(i_{\underline{j}})\left| e_{i}^{[j+1]}\right\rangle =0  
$$
and
$$
p_{j+1}(i_{\underline{j}},i) = \left\langle e_{i}^{[j+1]}\right|\rho_{j+1} (i_{\underline{j}})\left| e_{i}^{[j+1]}\right\rangle = \frac{1}{k_1\cdot k_2\dots \cdot k_{j+1}}
$$
for all $i= 1,\dots , k_{j+1}$, where 

$$
M_{j+1}(i_{\underline{j}}) = 
\left\langle e^{[\underline{j}]}_{i_{\underline{j}}} \right|M_{\underline{j+1}}\left|e^{[\underline{j}]}_{i_{\underline{j}}}\right\rangle ,\quad 
\rho_{j+1} (i_{\underline{j}})= 
\left\langle e^{[\underline{j}]}_{i_{\underline{j}}} \right|\rho_{\underline{j+1}}\left|e^{[\underline{j}]}_{i_{\underline{j}}}\right\rangle,\quad {\rm and}\quad
\left|e^{[\underline{j}]}_{i_{\underline{j}}}\right\rangle = 
\left|e^{[1]}_{i_1}\right\rangle \otimes \dots \otimes
\left|e^{[j]}_{i_j}\right\rangle
$$
Note that the second condition means that for each $j$ the outcome $i_j$ is independent of the others and has equal probabilities $1/k_j$.

After $n$ steps we have defined in an adpative fashion a product measurement basis 
$$
\left|
e^{[\underline{n}]}_{i_{\underline{n}}}
\right\rangle = \left|e^{[1]}_{i_1}\right\rangle \otimes \dots \otimes
\left|e^{[n]}_{i_n}\right\rangle
$$
and one can verify that such a measurement satisfies the general condition \eqref{eq:condition.Jiang}.

\section{Discrete quantum Markov chains and the output QFI}
\label{sec:Markov}

In the input-output formalism the dynamics of a discrete-time quantum open system $\mathcal{H}_s\cong\mathbb{C}^d$ is modeled by successive unitary interactions with independent `noise units', identically prepared in a state $|\chi\rangle\in \mathcal{H}_u\cong\mathbb{C}^k$. This can be pictured as a conveyor belt where the incoming `noise units' constitute the input, while the outgoing `noise units' make up the output of the process, cf. Figure \ref{Fig:i-o}. If $|\phi\rangle\in \mathcal{H}_s$ is the initial state of the system, and $U$ is the unitary on 
$\mathcal{H}_s\otimes \mathcal{H}_u$ describing the interaction between system and a noise unit, then the state of the system and output after $n$ time units is
\begin{equation}\label{eq:total.state}
|\Psi(n)\rangle = U(n) |\phi\otimes \chi^{\otimes n}\rangle = 
U^{(n)}\cdot\dots \cdot U^{(2)}\cdot U^{(1)}|\phi\otimes \chi^{\otimes n}\rangle\in 
\mathcal{H}_s\otimes \mathcal{H}_u^{\otimes n}
\end{equation}
where $U^{(i)}$ is the unitary acting on the system and the $i$-th noise unit.
From equation \eqref{eq:total.state} we find that the reduced state of the system at time $n$ is given by 
$$
\rho(n) := 
{\rm Tr}_{\rm out}
(|\Psi(n)\rangle \langle \Psi(n)|) = 
T^n(\rho_{\rm in}), \qquad \rho_{\rm in} = |\phi\rangle\langle \phi|,
$$
where the partial trace is taken over the output (noise units), and 
$
T:\mathcal{T}_1(\mathcal{H}_s)\to\mathcal{T}_1(\mathcal{H}_s)
$ 
is the Markov transition operator 
$$
T: \rho \mapsto {\rm Tr}_{u}(U (\rho \otimes \tau ) U^*),\qquad \tau := |\chi\rangle \langle \chi|.
$$
Fixing an orthonormal basis $\{|1\rangle, \dots ,  |k\rangle\}$ in 
$\mathcal{H}_u
$, we can express the system-output state as a matrix product state
\begin{equation}\label{eq.Psi_n}
|\Psi(n)\rangle = 
\sum_{i_1,\dots , i_n =1}^k 
K_{i_n}\dots K_{i_1} |\phi\rangle \otimes |i_{1}\rangle\otimes \dots\otimes |i_{n}\rangle
\end{equation}
where $K_i= \langle i|U|\chi\rangle$ are the Kraus operators of $T$, so that $T(\rho) = \sum_i K_i \rho K_i^*$.

Now, let us assume that the dynamics depends on a parameter $\theta\in \mathbb{R}$ which we would like to estimate, so that 
$U= U_\theta$ and $|\Psi(n)\rangle=|\Psi_\theta(n)\rangle$. In the input-output formalism it is usually assumed that the experimenter can measure the output (noise units after the interaction) but may not have access to the system. In this case the relevant quantum statistical model is that of the (mixed) output state given by 
$$
\rho^{\rm out}_\theta(n)
={\rm Tr}_s 
(|\Psi_\theta(n)\rangle \langle \Psi_\theta(n)|).
$$

The problem of estimating $\theta$ in this formulation has been investigate in both the discrete time 
\citep{Guta2011,GutaKuikas} and the continuous time \citep{GutaCatanaBouten,Molmer2014,GK17} settings. For our purposes, we summarise here the relevant results of \citep{GutaKuikas}. We will assume that the Markov chain is \emph{primitive}, i.e. the transition operator $T$ has a unique full-rank steady state $\rho_{\rm ss}$ (i.e. $T(\rho_{\rm ss}) = \rho_{\rm ss}$) , and is aperiodic (i.e. the only eigenvalue of $T$ on the unit circle is $1$). In particular, for any initial state $\rho_{\rm in}$, the system converges to the stationary state 
$T^n(\rho_{\rm in})\to\rho_{\rm ss}$ in the large $n$ limit. Therefore, for the asymptotic analysis we can assume that the dynamics is in the \emph{stationary regime} and focus on the large time properties of the stationary output state. The following Theorem shows that the output QFI scales linearly with time and provides an explicit expression of the rate.
\begin{theorem}
\label{Th.QFI}
Consider a primitive discrete time Markov chain as described above, whose unitary depends smoothly on a one-dimensional parameter $\theta$, so that  $U= U_\theta$. The quantum Fisher information $F_\theta(n)$ of the output  state 
$\rho^{\rm out}_\theta(n)$ scales linearly with $n$ and its rate is equal to
\begin{equation}
    \label{eq:QFI.rate}
\lim_{n\to\infty } \frac{1}{n} F_\theta(n) = f_\theta=  
4\sum_{i=1}^k \left[ {\rm Tr}\left[\rho_{ss}\dot{K}_i^*\dot{K}_i\right]  + 2{\rm Tr} \left[{\rm Im} ( K_i  \rho_{ss}  \dot{K}_i^* ) \cdot\mathcal{R}( {\rm Im} \sum_j\dot{K}_{j}^* K_j )\right]\right]
\end{equation}
where $\mathcal{R}$ is the Moore-Penrose inverse of ${\rm Id}- T_\theta$.
\end{theorem}

Following standard quantum Cram\'{e}r-Rao theory \citep{Helstrom1976,Holevo,BraunsteinCaves}, the theorem implies that the variance of any (unbiased) output-based estimator is bounded from below by $n^{-1}/f(\theta)$ for large $n$. A more in depth analysis \citep{GutaKuikas} shows that the output model satisfies the property of \emph{local asymptotic normality} which pertains to a certain quantum Gaussian approximation of the output state and implies that there exists an estimator $\hat{\theta}_n$ which achieves the CR bound asymptotically and has normally distributed errors:
$$
\sqrt{n}\left(\hat{\theta}_n -\theta \right) \longrightarrow N(0, f(\theta)^{-1})
$$
where the convergence is in distribution to a normal variable with variance $f(\theta)^{-1}$. Below, we will make use of an extension of Theorem \ref{Th.QFI} which shows that the same result holds for rank-deficient stationary states of ergodic chains, and in particular for pure states \citep{GutaKiukas3}.


Having identified the output QFI rate, we would like to investigate measurement schemes which can provide good accuracy for estimating the parameter $\theta$. As noted before, the QCRB can be achieved by measuring the SLD of the statistical model. However, the SLD of the output state is generally a complicated operator whose measurement requires collective operations on the noise units. On the other hand, one can consider separate measurements of the same observable on the different noise units and system. The average statistic may provide an efficient estimator and its (asymptotic) Fisher information can be computed explicitly \citep{GutaCatanaBouten}. However, such measurements are in general not optimal. Here would would like to ask the more fundamental question: is it possible to achieve the QCRB using simpler `local' manipulation of the output units which involve operations on single, rather than multiple units.

\section{Output post-processing using quantum coherent absorber}
\label{sec:pure.stationary}


In this section we introduce a key tool which will allow us to recast the estimation problem for a general primitive Markov chain ( which typically has a mixed stationary state) into one concerning a Markov chain with a `doubled-up' system having a \emph{pure stationary state}. The construction is a discrete-time adaptation on the concept of \emph{coherent quantum absorber} introduced in \citep{Preskill} for continuous-time dynamics. 
In section \ref{sec:adaptive.measurement} we then show how the absorber can be used to compute an adaptive, separable measurement in a simple recursive algorithm.  


Consider the input-output system of section \ref{sec:Markov} characterised by a 
unitary $U$ on $\mathcal{H}_s\otimes\mathcal{H}_u$. We now modify the setup as illustrated in Figure \ref{Fig:adapt-meas} by inserting an additional physical $d$-dimensional 
system $\mathcal{H}_a$ called absorber which interacts with each of the noise units via a fixed unitary 
$V$ on $\mathcal{H}_a\otimes \mathcal{H}_u$, applied immediately after $U$. This can be seen as a type of quantum post-processing of the output prior to the measurement. The original system and the absorber can be considered a single open system with space  $\mathcal{H}_s\otimes \mathcal{H}_a$ which interacts with the same conveyor belt of noise units via the unitary
$
W_\theta= V\cdot U
$
where $V, U$ are now understood as the ampliations of the unitaries to the tensor product $\mathcal{H}_s \otimes \mathcal{H}_a\otimes \mathcal{H}_u$. The following lemma shows that for certain choices of $V$ the auxiliary system forms 
a pure stationary state together with the original one, and the noise units pass unperturbed from input to output. This explains the `absorber' terminology, which was originally introduced in the context of continuous-time input-output dynamics \citep{Preskill}. 

\begin{lemma}\label{lemma.absorber}
Any given primitive Markov with unitary $U$ can be extended to a quantum Markov chain including an absorber with unitary $V$, such that the doubled-up system has a pure stationary state $|\tilde{\psi}\rangle\in \mathcal{H}_s\otimes \mathcal{H}_a$ and 
$$
W :|\tilde{\psi}\rangle \otimes |\chi\rangle \mapsto 
|\tilde{\psi}\rangle \otimes |\chi\rangle, \qquad 
W= VU.
$$
In particular, if the initial state of the doubled-up system is $|\tilde{\psi}\rangle$, then the $n$-steps output state of the doubled-up system is identical to the input state $|\chi\rangle^{\otimes n}$. 
\end{lemma}

\noindent
\emph{Proof.} 
Let $\rho_{\rm ss} = 
\sum_{i} \lambda_i |f_i\rangle \langle f_i|$ be the spectral decomposition of the stationary state of the original system with unitary $U$.  We construct the purification 
$$
|\tilde{\psi}\rangle =\sum \sqrt{\lambda_i} |f_i\rangle \otimes |f_i\rangle \in \mathcal{H}_s\otimes\mathcal{H}_a
$$
which will play the role of stationary state of the extended system. Let $|\phi\rangle := 
U |\tilde{\psi}\otimes \chi\rangle \in \mathcal{H}_s\otimes \mathcal{H}_a\otimes \mathcal{H}_u $ be the state after applying $U$. We therefore look for unitary $V$ on $\mathcal{H}_a\otimes \mathcal{H}_u$ (ampliated by identity on $\mathcal{H}_s$) such that $V$ reverts the action of $U$
$$
V :|\phi\rangle \mapsto |\tilde{\psi }\otimes \chi\rangle.
$$

Since $ {\rm Tr}_a (|\tilde{\psi}\rangle \langle\tilde{\psi}|) = \rho_{\rm ss}$ this means that the reduced state of the system after applying $U$ is still the stationary state, so that
$$
|\phi\rangle = 
\sum_{i} \sqrt{\lambda_i} |f_i\rangle \otimes |g_i\rangle
$$
where $|g_i\rangle$ are mutually 
orthogonal unit vectors in $\mathcal{H}_a\otimes \mathcal{H}_u$. We now choose a unitary $V$ such that 
$V|g_i\rangle= |f_i\otimes \chi\rangle$, for all $i$, which is always possible due to orthogonality.

\qed

\section{Adaptive measurement algorithm}
\label{sec:adaptive.measurement}

In this section we describe our adaptive output measurement protocol for estimating an unknown one-dimensional dynamical parameter $\theta$ of a discrete time 
quantum Markov chain with unitary $U_\theta$, as described in section \ref{sec:Markov}. The protocol has two stages. In the first stage we use a small proportion (e.g. $\tilde{n} = n^{1-\epsilon}$, with $0<\epsilon\ll 1 $) of the output units in order to compute a preliminary ‘rough estimator' $\theta_0$ of the true parameter $\theta$ by performing a standard sequential measurement. This step is necessary in
any quantum estimation problem in which the optimal measurement depends on the unknown parameter \citep{GillMassar,DemkowiczGuta}, and will inform the second stage of the protocol. In the second stage we use $\theta_0$  to design a \emph{optimal} sequential measurement for $\theta=\theta_0$, i.e. one that achieves the output QFI at $\theta=\theta_0$. Since $\delta \theta = \theta -\theta_0 =O(n^{-1/2 +\epsilon})$  \citep{GutaCatanaBouten}, this implies that the procedure is asymptotically optimal for \emph{any} parameter value $\theta$, in that the classical Fisher information has the same linear scaling as the QFI $F_\theta(n)$. This stage has two key ingredients (cf. Figure \ref{Fig:adapt-meas}): a  quantum `post-processing’ operation on output units immediately after interacting with the system, followed by an adaptive projective measurement whose basis is computed according to a ‘measurement filter’ algorithm inspired by \citep{Zhou20}. We now describe the two steps in detail.

\emph{Quantum post-processing.} 
After the interaction $U_\theta$ with the system, the output units interact sequentially with a d-dimensional ancillary system, cf. Figure \ref{Fig:adapt-meas}. The interaction unitary $V_{\theta_0}$ is chosen such that the ancillary system is a coherent quantum absorber for $\theta=\theta_0$, see section \ref{sec:pure.stationary} and Lemma \ref{lemma.absorber} for construction. This means that the system plus absorber (s+a) can be regarded as a single $D= d^2$ dimensional open system with associated unitary $W_\theta = V_{\theta_0}U_\theta$, which has a pure stationary state at $\theta=\theta_0$ denoted $|\psi \rangle \in \mathbb{C}^D$, and whose output state is identical to the input. The general estimation problem for $U_\theta$ has been reduced to a special one for a doubled-up system with unitary $W_\theta$ which features a pure stationary state at $\theta=\theta_0$. 
\begin{remark}\label{remark}
Since the absorber transformation $V_{\theta_0}$ does not depend on $\theta$ and is applied after $U_\theta$, the overall effect over an $n$ steps interval is to rotate the absorber plus output state by a fixed unitary $V^{(n)}_{\theta_0}\dots V^{(1)}_{\theta_0}$. This means that the total QFI does not change by introducing the absorber.    
\end{remark}



\emph{Adaptive measurement algorithm.}
We will assume for simplicity that the initial state of $s+a$ is the stationary state $|\psi\rangle$ such that the full (s+a)-output state is 
$|\Psi_\theta(n)\rangle  = W_\theta (n) |\psi\otimes \chi^{\otimes n}\rangle$ as defined in 
\eqref{eq:total.state}. One could obtain similar results for different initial states by simply waiting long enought for the system and absorber to converge to the stationary state $|\psi\rangle$. 
In principle we could now apply the algorithm described in section \ref{sec:Jiang} to construct and adaptive measurement whose classical Fisher information is equal to the system-output QFI. In fact we could have done this without using the absorber. However, this procedure has some drawbacks.  Indeed, in order to compute the measurement bases one needs to work with large dimensional spaces which becomes unfeasible in an asymptotic setting. Secondly, it is not clear a priori whether the output units can be measured immediately after the interaction with the system, and whether the measurements depend on the length of the output (sample size). In addition, the procedure requires a final measurement on the system, which may be impractical in the context of input-output dynamics. We will show that all these issues can be addressed by taking into account the Markovian structure of our model, and exploiting the pure stationary state property. Let us denote 
$W= W_{\theta_0}$, $\dot{W}= \left. \frac{dW}{d\theta} \right|_{\theta_0} $ and  
$$
A_1 =  M^{(1)}= \dot{W}|\psi\otimes \chi\rangle  \langle \psi\otimes \chi | W^* - W|\psi\otimes \chi\rangle  \langle \psi\otimes \chi | \dot{W}^* =
\dot{W} P_{\psi\otimes \chi} 
- P_{\psi\otimes \chi}   \dot{W}^*,
$$
and
$$
B_1 = {\rm Tr}_{s+a} A_1 = K P_\chi -P_\chi K^*, \qquad {\rm with}\qquad
K= \langle \psi |\dot{W}|\psi\rangle. 
$$
In Appendix \ref{app:filter.derivation} we show that the adaptive measurement of \citep{Zhou20} reduces to the following iterative algorithm which is  conceptually similar to quantum state filtering \citep{Belavkin94,BvHJ07}, and involves individual measurements on the output units immediately after the interaction with the system, and computations with operators on $\mathbb{C}^D\otimes \mathbb{C}^k$ at each step.

\vspace{2mm}

\noindent
{\bf Initialisation step (j=1).}
    The first measurement basis $\left\{\ket{e^{[1]}_{i}} \right\}$ in $\mathbb{C}^k$ is chosen such that the following conditions are fulfilled: 
    $$
       \left\langle e^{[1]}_{i} \right| 
         B_1 
        \left| e^{[1]}_{i} \right\rangle  
        =0, 
        \qquad {\rm and }\qquad 
        \left|\braket{e^{[1]}_{i}|\chi}\right|^2=\frac{1}{k}, \qquad {\rm for~all~} i=1\dots, k.
    $$
    The first noise unit is measured in this basis and the outcome $X_1= i_1$ is obtained. 
    The filter at time $j=1$ is defined as the (trace zero) s+a operator
    $$
    \Pi_1 = 
    \left\langle e^{[1]}_{i_1} \right| 
A_1
        \left| e^{[1]}_{i_1} \right\rangle.
    $$
    
    

   \vspace{2mm}
   
 {\bf Iterative step.} The following step is iterated for $j=2,\dots , n$. Given the filter operator 
    $\Pi_{j-1}$ of the previous  step, we define
    $$
    A_j = \frac{1}{D^{j-1}} A_1 + W \left(  \Pi_{j-1} \otimes 
    P_\chi \right)  W^* , \qquad B_j = {\rm Tr}_{s+a} A_j
    $$
The $j$-th measurement basis $\left\{\ket{e^{[j]}_i} \right\}$ is chosen to fulfill the conditions
    \begin{equation}
    \label{eq:conditions.basis}
        \left\langle e^{[j]}_{i} \right| 
         B_j 
        \left| e^{[j]}_{i} \right\rangle
        =0 ,
        \qquad {\rm and}\qquad 
        \left|\braket{e^{[j]}_{i}|\chi}\right|^2=\frac{1}{k}, \qquad {\rm for~all~}i=1,\dots k.
    \end{equation}
    We measure the $j$-th noise unit in the basis 
    $\left\{ |e^{[j]}_{i}\rangle \right\}$ and obtain the result 
    $X_j = i_j$. The filter at time $j$ is updated to 
    $$
    \Pi_j = \left\langle e^{[j]}_{i_j} \right| 
         A_j
        \left| e^{[j]}_{i_j} \right\rangle.
    $$

{\bf Final s+a measurement.}
This is an optional step which involves a final joint measurement on system and absorber. The basis $\left\{ |e^{[s+a]}_i\rangle \right\}_{i=1}^D$ is  determined by the following conditions
$$
\left\langle e^{[s+a]}_{i} \right| \Pi_n \left| e^{[s+a]}_i\right\rangle = 0, 
\qquad p(i|i_1,\dots , i_n) = \left|\langle e^{[s+a]}_i|\psi\rangle \right|^2 =1/D.
$$      
 The system and absorber is measured in this basis and the outcome $X=i_0$ is obtained.   

The output measurement record $\{i_1, \dots , i_n, i_{0}\}$ is collected and used for estimating the parameter $\theta$. The likelihood function is given by
    \begin{equation}\label{eq:likelihood}
    p_\theta(i_1, \dots i_n, i_0) = 
    \left| \left\langle e^{[s+a]}_{i_0} \otimes e^{[1]}_{i_1}\otimes \dots \otimes e^{[n]}_{i_n} \left. \right|  \Psi_\theta(n)\right\rangle\right|^2.
\end{equation}
For later use, we denote by $p_\theta(i_1, \dots i_n)$ the marginal distribution of the output measurement record only. 
    

\section{Fisher informations considerations}

In this section we investigate the relationship between the classical Fisher information (CFI) of the output adaptive measurement process and the system-output QFI. We prove that both scale with the same rate and the latter may be larger than the former by at most a constant, independent of time. 

We also provide an 
expression of the CFI of sequential (adaptive or standard) output measurements, which is amenable to estimation by sampling. This tool will be used to confirm the optimality of our adaptive algorithm in numerical simulations. 

\subsection{Achievability of the QFI with adaptive output measurements}

The adaptive measurement scheme described in section \ref{sec:adaptive.measurement} insures the CFI of the full measurement (output and s+a) is equal to the QFI of the full pure state model
\begin{equation}\label{eqn:optimal.FI}
    I^{({\rm s+a+o})}_{\theta_0} (n) = 
    F^{({\rm s+a+o})}_{\theta_0}(n) = 
    F^{({\rm s+o})}_{\theta_0}(n)
\end{equation}
where the last equality follows from the fact that the absorber acts as an additional rotation which does not change the QFI, cf. Remark \ref{remark}.
However, in certain physical implementations the system may not be accessible for measurements, so the more interesting scenario is that in which only the output state is measured. 
In this case the CFI will generally be strictly smaller that the QFI, and the question is whether by measuring only the output we incur a significant loss of information. In proposition \ref{prop:CFI-QFI} we show that this is not the case: the difference between the QFI and the output CFI is bounded by a constant, so for large times the loss of information is negligible compared to both QFI and output CFI, which scale linearly with time. In section \ref{sec:Simulations} we will illustrate the result on a specific model. 

\begin{proposition}
\label{prop:CFI-QFI}
Consider the setup described in section \ref{sec:adaptive.measurement},  and let $F(n)$ be the system-absober-output QFI, and $I^{\rm (o)}(n)$ be the output CFI for the optimal adaptive measurement, at $\theta=\theta_0$. Then $F(n)- I^{\rm (o)}(n) <c$ for all $n$ where $c$ is a constant depending only on the model $U_\theta$. Consequently, 
$$
\lim_{n\to \infty}
\frac{1}{n}I^{\rm (o)}(n) = \lim_{n\to \infty}
\frac{1}{n}F(n) = f>0
$$
where $f =f_{\theta_0}$ is the QFI rate \eqref{eq:QFI.rate}.
 \end{proposition}

\subsection{Computing the classical Fisher information of the output}
\label{sec:classical_Fisher}

The \emph{classical Fisher information} of the output measurement process at $\theta_0$ is
$$
I^{{\rm (o)}}(n) =\mathbb{E}_{\theta_0}
\left( 
\frac{d\log p_{\theta}}{d\theta}
\right)^2
=\sum_{i_1,\dots, i_n} p_{\theta_0}(i_1, \dots i_n)^{-1} \left(\left.\frac{d p_{\theta}(i_1,\dots , i_n)}{d\theta} \right|_{\theta_0}\right)^2
$$
where the sum runs over indices such that $p_{\theta}(i_1,\dots , i_n)>0$. In general 
$$
I^{{\rm (o)}}(n)\leq F^{{\rm (o)}}(n) \leq F^{({\rm s+o})}(n)
$$
where the successive upper bounds are output and system-output QFIs respectively. 

In our simulation study we will be interested to study to what extent these bounds are saturated in the adaptive and non-adaptive scenarios, and in particular, to verify the prediction of Proposition \ref{prop:CFI-QFI}. Since the classical Fisher information is difficult to compute for long trajectories, we will recast it as an expectation which can be estimated by sampling measurement trajectories. In Lemma \ref{lemma:CFI-simulations} below, we will use the fact that at $\theta=\theta_0$ the vector $|\psi\rangle$ is the stationary state, and therefore 
\begin{equation}\label{eqn:kraus.constants}
    K^{[j]}_{i}|\psi\rangle = c^{[j]}_i |\psi\rangle
\end{equation}
for \emph{any} Kraus decomposition $K_i^{[j]} = \langle e_j^{[i]} |W|\chi\rangle$ (for simplicity we use the same notation for s+a Kraus operators as in section \ref{sec:Markov}). The proof of Lemma \ref{lemma:CFI-simulations} can be found in Appendix \ref{app:proof_lemma_simulations}.

\begin{lemma}\label{lemma:CFI-simulations}
Consider the setup described in section \ref{sec:adaptive.measurement}. The output CFI at 
$\theta= \theta_0$ is given by 
\begin{equation}
I^{{\rm (o)}}(n) =  \mathbb{E}_{\theta_0} (f^2)=
\sum_{i_1,\dots, i_n} 
p_{\theta_0}(i_1,\dots, i_n) f^2(i_1,\dots, i_n) 
\label{eq:classical.Fisher}
\end{equation}
where $f$ is the function 
\begin{equation*}
f(i_1,\dots , i_n) = 2{\rm Re}\sum_{j=1}^n 
\frac{\langle \psi|K^{[n]}_{i_n}\dots K^{[j+1]}_{i_{j+1}} \dot{K}^{[j]}_{i_j}|\psi\rangle }{c^{[j]}_{i_{j}} \dots c^{[n]}_{i_n}}
\end{equation*}
and the constants $c^{[j]}_{i_j}$ are defined by equation \ref{eqn:kraus.constants}. In particular, $I^{{\rm (o)}}(n)$ can be estimated by computing the empirical average of  $f^2$ over sampled trajectories.
\end{lemma}

We now consider the case where a (projective) measurement 
$\{P^{({\rm s+a})}_{i}\}$ is performed on s+a, after obtaining the output measurement record $(i_1,\dots , i_n)$. We denote the additional outcome by  $i_{0}$. The CFI of the full process is
$$
I^{({\rm s+a+o})}(n) = \mathbb{E}_{\theta_0}
\left( \left.
\frac{d\log p_\theta}{d\theta}\right|_{\theta_0}
\right)^2
$$
where 
$$
p_\theta (i_1,\dots , i_n, i_{0})=
\left\| P^{({\rm s+a})}_{i_{0}} K^{[n]}_{i_n}\dots K^{[1]}_{i_{1}} \psi\right\|^2
$$ 
is the likelihood function of a trajectory augmented by the system measurement outcome $i_{0}$. The relevant upper bound in this case is 
\begin{equation}
\label{eq:CFI-QFI.bound.so}
I^{({\rm s+a+o})} (n)\leq F^{({\rm s+a+o})} (n) = F^{({\rm s+o})}(n).
\end{equation}
A similar computation to that of Lemma \ref{lemma:CFI-simulations} gives the system-output classical Fisher information 
$$
I^{({\rm s+a+o})}(n) = \mathbb{E}_{\theta_0} (\tilde{f}^2)
$$ 
where 
$\tilde{f}$ is the function
\begin{eqnarray*}
\tilde{f} (i_1,\dots, i_n, i_{0}) &= &
2\,{\rm Re}\sum_{j=1}^n 
\frac{\langle \psi| P^{({\rm s})}_{i_{0}}K^{[n]}_{i_n}\dots K^{[j+1]}_{i_{j+1}} \dot{K}^{[j]}_{i_j}|\psi\rangle }{c^{[j]}_{i_{j}} \dots c^{[n]}_{i_n}}\\
\end{eqnarray*}

For fixed (non-adaptive) measurements, the bound \eqref{eq:CFI-QFI.bound.so} is generally not saturated except for special models 
(e.g. if the state coefficients in the measurement basis are real for all $\theta$). In contrast, the system-absorber-output classical Fisher information for adaptive measurements is equal to the QFI thanks to the optimality of the adaptive measurement procedure \ref{eqn:optimal.FI}. This is confirmed by our simulation study which also investigates the performance of the fixed measurement scenario.

\section{Numerical simulations}
\label{sec:Simulations}

We now test the key properties of the adaptive measurement scheme developed in section \ref{sec:adaptive.measurement}, in two separate numerical investigations. 

The first investigation described in subsections \ref{sec:Markov.model} and \ref{sec:simplified.study} employs a simplified Markov model which bypasses stage-one of the scheme (computing a rough estimator $\theta_0$) and simulates data at $\theta=\theta_0$.
This allows us to directly study the performance of the algorithm itself (stage two), rather than that of the combination of the two stages. The second simplification of this investigation is that we choose a system which has a pure stationary state at $\theta_0$; this means that no absorber is required, so the system can be seen as a surrogate for the system+absorber in the general scheme. The reason for this is mainly practical, as it allows us to use a two-dimensional system while system+absorber would have dimension at least four.

The second numerical investigation consists of a full simulation study including the use of the coherent absorber and the two stage estimation procedure, and its results are presented in subsection \ref{sec:full.simulation}. Here we can see the overall performance of the estimation method, but it is harder to estimate the  Fisher information of the measurement process and to separate the contribution of the two stages in the overall estimation error.

\subsection{Simplified Markov model for the first numerical investigation}
\label{sec:Markov.model}
We consider a dynamical model consisting of a two-dimensional system coupled via a unitary $U_\theta$ to two dimensional noise units in state $|\chi\rangle = |0\rangle$. The input state and the unitary are designed such that the stationary state at $\theta_0=0$ is 
$|\psi \rangle =|0 \rangle$. 
Since the input is prepared in a fixed state, we only need to define the action of $U_\theta$ on the basis vectors $|0\rangle \otimes |0\rangle$ and 
$|1\rangle \otimes |0\rangle$. The following choice has unknown parameter $\theta$ and two known parameters $\lambda$ and $\phi$
\begin{align}
    U_\theta : &\ket{00} \longrightarrow \cos(\theta) \sqrt{1-\theta^2} \ket{00} + i \sin(\theta) \sqrt{1-\theta^2} \ket{10} + \theta \ket{11}, \nonumber \\
   U_\theta : &\ket{10} \longrightarrow i \sin(\theta) \sqrt{1-\lambda} \ket{00} + \cos(\theta) \sqrt{1-\lambda} \ket{10} + \sqrt{\lambda} e^{i \phi} \ket{01}.\label{eqn:SimulationUnitary}
\end{align}
A non-zero value of the phase parameter $\phi$  ensures that the system-output state does not have real coefficients in the standard basis, in which case the standard basis measurement would be optimal. Note that at $\theta_0$ the dynamics provides a simple model for `photon-decay' with decay parameter $\lambda$.



We compare two measurements scenarios. In the first, non-adaptive scenario, the noise units are measured in a fixed orthonormal basis $\{|f_0 \rangle = (|0\rangle+ |1\rangle)/\sqrt{2}, |f_1 \rangle = (|0\rangle-  |1\rangle)/\sqrt{2}\} $ while in the second scenario they are measured adaptively following the algorithm described in section \ref{sec:adaptive.measurement}. 
In both cases, given the measurement record $\{i_1, \dots ,  i_n\}\in \{0,1\}^n $, the conditional state of the system is 
\begin{equation}
\label{eq:state.filter}
|\psi_n(i_1,\dots i_n) \rangle= \frac{K^{[n]}_{i_n} \dots K^{[1]}_{i_1}|\psi \rangle }{\|K^{[n]}_{i_n} \dots K^{[1]}_{i_1} \psi\| }
\end{equation}
where the Kraus operators are 
$\{ K_0= \langle f_0|U|\chi\rangle,  K_1 =  \langle f_1|U|\chi\rangle\} $ in the first scenario, and  $K^{[j]}_i=\langle e^{[j]}_{i}|U|\chi\rangle$ in the second one. 
The likelihood of the output measurement trajectory is 
\begin{equation}\label{eq:likelihood_2}
p_\theta(i_1, \dots i_n)  
= 
\| K^{[n]}_{i_n}\dots K^{[1]}_{i_1}\psi \|^2.
\end{equation}
Recall that the optimal measurement basis needs to satisfy the conditions \eqref{eq:conditions.basis} in section \ref{sec:adaptive.measurement}. 
For two dimensional noise units the computation reduces to the following scheme. We express the traceless, anti-Hermitian matrix $B_j$ defined in section \ref{sec:adaptive.measurement} as $B_j = i \vec{r}^{\ [j]} \cdot \vec{\sigma}$ where 
$\vec{r}^{\ [j]} = (r^{[j]}_x,r^{[j]}_y,0)$ is its  Bloch vector, and similarly, we let $\pm \vec{s}^{\ [j]}$ be the Bloch vectors of the basis vectors $\{|e_1^{[j]}\rangle ,|e_2^{[j]}\rangle\} $ satisfying the conditions \eqref{eq:conditions.basis}. Then one finds that the conditions \eqref{eq:conditions.basis} are satisfied if $\vec{s}^{\ [j]} $ is taken to be $\vec{s}^{\ [j]} = (r^{[j]}_y, -r^{[j]}_x,0)$.

We study both the scenario where system of interest is measured after obtaining the output trajectory, as well as the one where only the output measurement is considered. In the former  case, the algorithm guarantees that the classical Fisher information of the full measurement record is equal to the QFI of the system-output state. This claim is verified numerically by comparing the estimated classical Fisher information computed using the method described in section \ref{sec:classical_Fisher} with the \emph{quantum} Fisher information of the system-output state. In the latter scenario, Proposition \ref{prop:CFI-QFI} insures that the loss of information compared to a `full measurement' is bounded by a constant which does not depend on time.


As shown in Theorem \ref{Th.QFI}, the system-output QFI scales linearly with time, i.e.
$$
F^{({\rm s+o})}_\theta (n) = n f_\theta + o(n)
$$
where the QFI \emph{rate} is given by the equation \eqref{eq:QFI.rate} and the $o(n)$ term depends on the specific system parameters. Applying this to our model we obtain
\begin{equation}
\label{eq:QFI.rate.model}
f_{\theta_0} = \frac{8}{1-\sqrt{1-\lambda}} .
\end{equation}
However, it turns out that for this specific model and parameter value, the QFI can be computed explicitly for \emph{any} fixed $n$. The proofs of \eqref{eq:QFI.rate.model} and of the following lemma can be found in Appendix \ref{app:QFI-finite-time}.

\begin{lemma}
\label{lemma:QFI-finite-time}
The system-output QFI at 
$\theta =\theta_0$ is given by the formula
\begin{eqnarray}
F^{({\rm s+o})}_\theta(n) &=& 
 =\frac{8n}{1-a}+
4\left[ 
 \frac{2(a^2 -a^n)}{(1-a)^2} 
-2 \frac{b^2 (a - a^{n})}{(1-a)^3} +  \frac{2(a^2-a^{2n})}{b^2} \right] 
\end{eqnarray}
where $a = \sqrt{1-\lambda}$ and $b=\sqrt{\lambda}$.  In particular, the leading term in $n$ is given by \eqref{eq:QFI.rate.model} while the remaining terms are bounded.

\end{lemma}


%


\subsection{Simulation studies for the simplified model}\label{sec:simplified.study}
We now present the results of our first numerical investigation consisting of 3 simulation studies using the simplified Markov model described above.

The first simulation study focuses on the comparison between the different notions of Fisher information: the system-output QFI, the CFI of the output trajectory in the non-adapted and adapted scenarios, and the CFI of the  system-output measurement process in the adapted measurement scenario. The QFI is computed using the formula in Lemma \ref{lemma:QFI-finite-time} while the CFIs are estimated by sampling using the expression in Lemma \ref{lemma:CFI-simulations}. 

\begin{figure}[h]
    \centering
    \includegraphics[width=8cm]{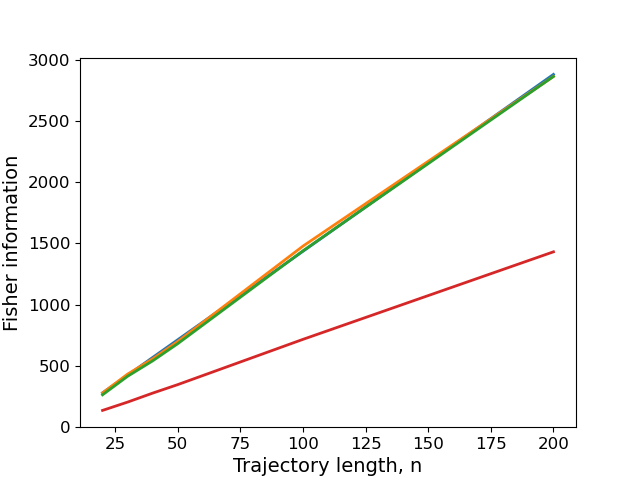}
    \caption{Fisher informations as function of output lenght $n$: quantum Fisher information (blue), classical Fisher information for the adaptive measurement with/without system measurement (orange/green), classical Fisher information for a regular (non-adaptive) measurement.}
    \label{Fig:Fisher-infos}
\end{figure}

The results are illustrated in Figure \ref{Fig:Fisher-infos} where the different informations are plotted as a function of time (trajectory length) $n$, for $\lambda=0.8,\phi= \pi/4$. 
The simulation confirms the fact that the adaptive algorithm achieves the QFI when the system is measured together with the output, while the CFI of the output trajectory (without measuring the system) provides a close approximation which only differs by a constant factor. 
In contrast, the CFI of the standard measurement has a smaller rate of increase; additional numerical work shows that the CFI rate can be improved by optimising the basis of the standard measurement but it does not achieve the QFI.

\vspace{4mm}

\begin{figure}[h]
    \centering
    \includegraphics[width=7cm]{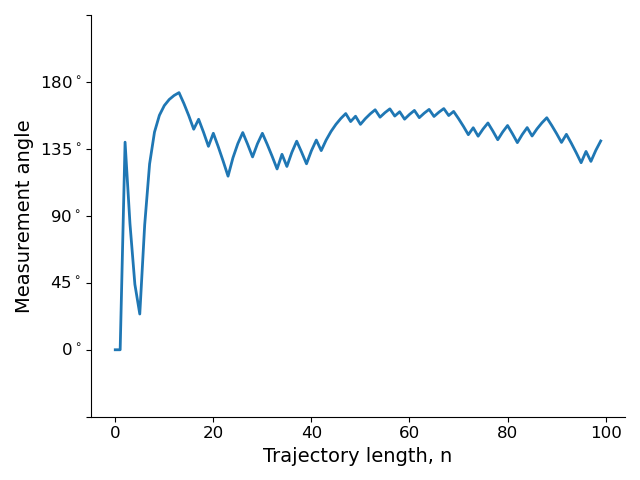}
    \includegraphics[width=7cm]{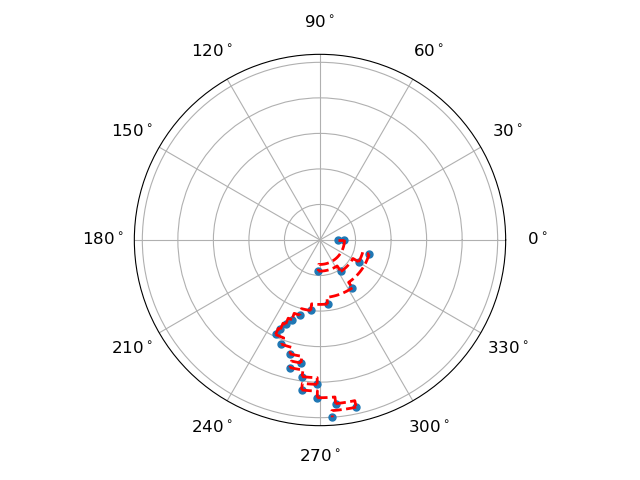}
    \caption{Adaptive measurement trajectory with basis angle $\varphi$ on left and $\varphi$ plotted on the circle on the right. }
    \label{Fig:angle_trajectory}
\end{figure}

The second simulation study focuses on the `measurement trajectory', i.e. the sequence of measurement settings produced in the adaptive measurement scenario. Since all measurement bases consist of vectors in the equatorial plane on the Bloch sphere, one can  parametrise each basis by the polar coordinate $\varphi$ of the basis vector 
$(|0\rangle \pm e^{i\varphi}|1\rangle)/\sqrt{2}$ for which $\varphi$ belongs to a specified interval of length $\pi$. This is illustrated in the left panel of Figure \ref{Fig:angle_trajectory}. This parametrisation has the disadvantage that it does not reproduce the topology of the space of measurements which is that of a circle, leading to some jumps in measurement angles appearing to be larger than the actual `distance' between measurement bases. To remedy this, in the right panel of Figure \ref{Fig:angle_trajectory} we plot $2\varphi$ on circles of  radius increasing linearly with time.
We note that the initial steps of the trajectory show large variations which then `stabilise' around a certain range of values. This has to do with the fact that the initial angle can be chosen arbitrarily in this model as $B_1 =0$, cf. section \ref{sec:adaptive.measurement}. 
Understanding the nature of this stochastic process remains an interesting topic of future research.


\vspace{4mm}

\begin{figure}[h]
    \centering
    \includegraphics[width=8cm]{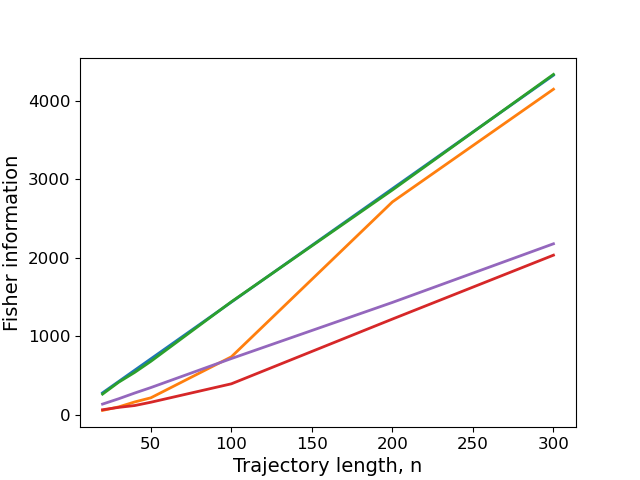}
    \caption{Comparison of different Fisher informations of the output state as function of trajectory length $n$, obtained by averaging over $N=10000$ trajectories, at $\lambda = 0.8, \varphi =\pi/4.$ The quantum Fisher information (QFI) (blue) and the classical Fisher information (CFI) of adapted measurement (green) completely overlap due to optimality. The inverse MLE error of the adapted measurement (orange) approaches the QFI for large $n$. Similarly, for the standard measurement, the inverse MLE error (red) approaches the CFI (purple) for large $n$.
    }
    \label{Fig:MLE_errors}
\end{figure}
\begin{figure}
    \centering
    \includegraphics[width=8cm]{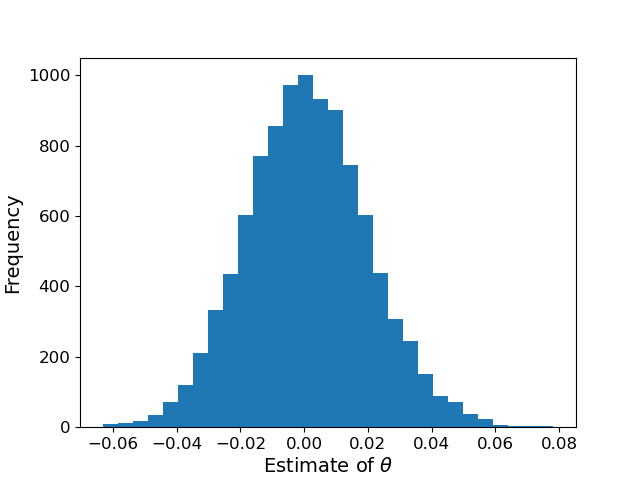}
    \caption{Histogram of the MLE distribution in the adaptive measurement scenario, with $n=200, \theta=0$ and  $10000$ samples, at $\lambda = 0.8, \varphi =\pi/4.$}
    \label{Fig:histogram}
\end{figure}

The third simulation study concerns the performance of the maximum likelihood estimator (MLE) in an adapted and non-adapted output measurement scenarios. The MLE is defined by
$$
\hat{\theta}_{n}:= \underset{\tau }{\arg\max} \,\,p_\tau(i_1, \dots i_n).
$$
where the likelihood $p_\tau(i_1, \dots i_n)$ is computed as in equation \eqref{eq:likelihood_2}. In numerics, we maximise the log-likelihood function which can be computed as the sum
$$
\log p_\tau(i_i,\dots i_n) = 
\sum_{j=1}^n \log \|K^{[j]}_{i_j} \psi (i_1,\dots i_{j-1}) \|^2
$$
where  $\psi (i_1,\dots i_{j-1})$ is the system's state conditional on the output trajectory (filter), cf. equation \eqref{eq:state.filter}. 
The MLE accuracy is quantified by the mean square error (MSE) $E(n)= \mathbb{E}_\theta (\hat{\theta}_n -\theta)^2$, which is estimated empirically by averaging over a number of simulations 
runs to obtain $\hat{E}(n)$. To verify that the MSE scales as $n^{-1}$ we plot the inverse empirical error 
$\hat{E}^{-1}(n)$ as a function of time. Figure \ref{Fig:MLE_errors} shows the inverse error for the adaptive and non-adaptive measurements together with the QFI (which is equal to the CFI of the adaptive measurement) and the CFI of the non-adaptive measurement. We note that for small values of $n$ the inverse error is significantly lower that the corresponding Fisher information, but it approaches the latter for larger values of $n$. This suggests that the MLE achieves the Cram\'{e}r-Rao bound asymptotically, which is not surprising since the MLE is known to be asymptotically optimal for independent samples as well as for certain classes of hidden Markov chains \citep{Lehmann,Bickel}. However, proving its optimality for the adaptive measurement process remains an open problem. In addition to the mean square error, we looked at the distribution of the MLE. Figure \ref{Fig:histogram} shows a histogram of the MLE based on $N=10000$ simulations with $n=200$, and it indicates that the MLE is approximately normally distributed. Based on this, it is reasonable to conjecture that the MLE is an \emph{efficient estimator} \citep{Lehmann,Bickel} (i.e. has asymptotically normal distribution with variance equal to the inverse of the Fisher information).

\subsection{The second numerical investigation using the full adaptive protocol}\label{sec:full.simulation}
In this section we present the results of the second numerical investigation which implements the full estimation scheme proposed in this paper, including the preliminary estimation stage and the use of the coherent absorber. We use the same input-output model as described by the unitary in equation \eqref{eqn:SimulationUnitary}, but the data will be simulated at a true value of $\theta=0.2$ instead of $\theta =0$. While at $\theta=0$ the system has a pure stationary state and the coherent absorber is not needed, away from this value the stationary state is mixed and we will apply the full protocol described in section \ref{sec:pure.stationary}.

In the first stage of the adaptive estimation procedure we use a fixed proportion $q$ of the total sample size $n$ to obtain a preliminary estimator $\theta_0$ of $\theta$ by measuring each output unit in the standard basis. The parameter is estimated using the maximum likelihood method. In the second stage we apply the adaptive scheme with 
an absorber `tuned' to the parameter value $\theta_0$ (cf. section \ref{sec:pure.stationary}). The system and absorber are prepared in the pure stationary state at $\theta_0$, the dynamics is run for time $(1-q)n$ and the output is measured according to the adaptive measurement filter algorithm. The maximum likelihood estimator for the stage one \emph{and} two data is then computed. 

For comparison, we also run a non-adaptive scheme where the output is measured in the standard basis for the whole duration $n$, without using an absorber. The maximum likelihood estimator is again computed from the measurement data. In both experiments, the mean square error of the MLE is estimated by averaging over 1000 repetitions.

 Unlike the setup of the previous numerical study, the estimation of the classical Fisher information of the adaptive measurement process was too costly and is not included in the study. As a proxy for the classical Fisher information we plot the average value of the \emph{observed Fisher information} for each of the two simulations. For a given measurement run $(i_1,\dots i_n)$, the observed Fisher information is defined as the second derivative of the log-likelihood function evaluated at the maximum likelihood estimator $\hat{\theta}_n$:
$$
I_{\rm obs} (i_1,\dots i_n) = - 
\left.\frac{d^2 \log p_\theta(i_1,\dots i_n)}{d\theta^2}\right|_{\theta = \hat{\theta}_n}
$$
While a full theoretical justification of $I_{\rm obs}$ goes beyond the scope of this paper, we note that the use of the observed Fisher information for independent identically distributed data is well grounded in statistical methodology and is closely relate to the asymptotic normality property \citep{Efron}.

\begin{figure}
    \centering
    \includegraphics[width=7.3cm]{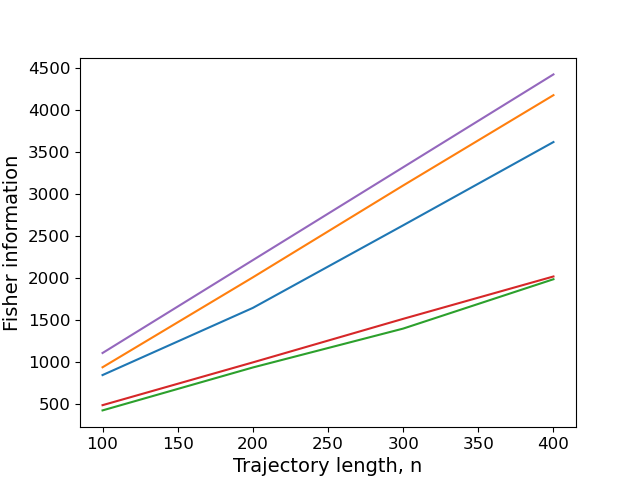}
    \includegraphics[width=7.3cm]{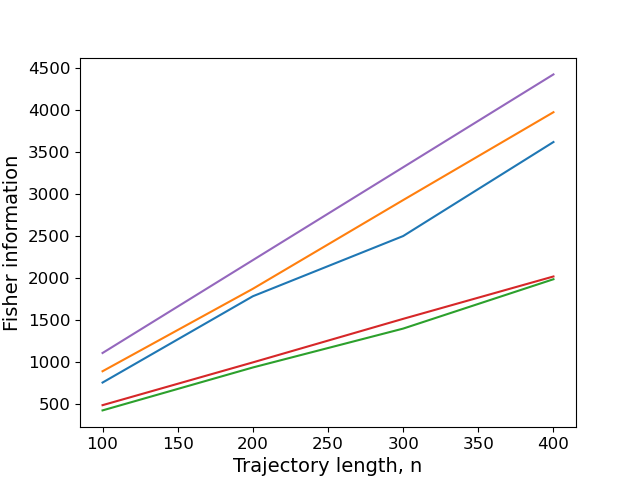}
    \caption{
    Comparison of different Fisher informations of the output state as function of trajectory length $n$, in the fully adaptive numerical investigation, for two proportions of samples used in the preliminary stage: $q=0.15$ (left panel) and $q=0.25$ (right panel). The results are obtained by averaging over $N=1000$ trajectories, at $\theta=0.2$ and $\lambda = 0.8, \varphi =\pi/4.$ We plot: the asymptotic QFI (purple line), observed Fisher information for adaptive measurement (orange line) and simple measurement (red line), inverse MSE of the MLE for adaptive measurement (blue line) and simple measurement (green line). }
    \label{fig.full.simulation}
\end{figure}

Figure \ref{fig.full.simulation} shows the results of the numerical experiments, for two values $q=0.15$ and $q=0.25$ of the proportion of samples used in the preliminary estimation stage. As before, we plot the inverse of the estimated mean square error for the simple measurement (green line) and the adaptive measurement (blue line). In addition we plot the leading contribution to the quantum Fisher information $n f(\theta)$ (purple line) computed using the results in Theorem \ref{Th.QFI}, which provides the asymptotic slope of the actual output quantum Fisher information. 
The average observed Fisher information is plotted for both the simple (red line) and adaptive (orange line) measurement setups. We note a good agreement between the inverse mean square error of the MLEs and the observed Fisher informations. For the adaptive measurements, the observed Fisher information also shows the same slope as the asymptotic Fisher information, as expected. We also note that in the adaptive measurement, the mean square error does not quite achieve the (observed) quantum Fisher information for the range of times we considered, although it gets closer to in the case $q=0.25$ (right panel). We speculate that this may be related to a number of factors such as the choice of $q$ for different sample sizes, the details of the actual measurement procedure implemented in practice (in contrast to the theoretical prescription) and even the implementation of the MLE. For instance, in practice it may be beneficial to implement several adaptive estimation stages where the preliminary estimator is gradually improved and used in tuning the absorber in the next stage. All these remain interesting questions which are worth investigating in more detail and on a case by case basis. However, these are somewhat separate issues from that of designing adaptive measurements that achieve the QFI, which was the main focus of this work.

\section{Conclusions and Outlook}

\label{sec:conclusions}

In this paper we developed an efficient iterative algorithm for optimal estimation of dynamical parameters of a discrete-time quantum Markov chain, using adaptive sequential measurements on the output. The algorithm builds on the general measurement scheme of \citep{Zhou20} which achieves the quantum Fisher information for pure state models of multipartite systems with one dimensional unknown parameters. However, unlike the scheme of \citep{Zhou20} which requires manipulations involving the full multipartite state, the proposed algorithm only involves computations on $d^2\cdot k$-dimensional systems where $d$ and $k$ are the dimensions of the open system and noise unit respectively. Therefore, the method can be readily applied to Markov parameter estimation for large output sizes. The algorithm exploits the Markovian structure of the dynamics to sequentially compute optimal measurement bases in terms of a single time-dependent `measurement filter' operator, which is updated in a way that is reminiscent of a state filter. One of the key ingredients of the proposed scheme is the use of a coherent quantum absorber \citep{Preskill} which reduces the estimation problem to one concerning a system with a pure stationary state. We considered both output and output-system measurements scenarios and we showed that while the former achieve the full quantum Fisher information, the latter is short of this by just a fixed constant, and in particular both have the same scaling with time. Our theoretical results are confirmed by numerical simulations using a simplified model related to the amplitude decay channel. We also presented results from `full simulation' study involving the use of the coherent absorber.

Our discrete-time procedure raises interesting questions about the the possibility 
to design realistic optimal sequential measurements in continuous-time dynamics. In principle the scheme can be applied to continuous time by using time-discretisation techniques \citep{Osborne,Horssen13}. Although we did not treat this in detail here, we can readily make several observations in the special case of a single Bosonic I-O channel. For small enough time intervals $\delta t$, the field can be approximated by a noise unit $\mathbb{C}^2$ with basis vectors representing the vacuum and a one-photon state respectively. The proposed measurements consist of projections whose corresponding Bloch vectors are in the equatorial plane; loosely speaking, this corresponds to an adaptive homodyne measurement with a time-dependent angle. Our preliminary investigations indicate that the behavior of the time-dependent angle ranges from deterministic evolution to a noisy stochastic process which does not appear to be of diffusive type. This raises the question whether the remaining freedom in choosing the measurement bases can be used to improve the adaptive algorithm and produce a more regular measurement processes. Indeed, the second defining condition for the measurement vectors can be relaxed, allowing for more general classes of optimal measurements, which may be more suitable for continuous-time direction. This will be the topic of a future investigation.

Another important open question concerns the extension to chains with mixed input states, or more multiple inputs, of which only some are observed. We speculate that for small departures from the current scheme, the algorithm will be quasi-optimal for some time interval but will be sub-optimal in the long time limit. In this case, restarting the evolution and measurement filter at regular intervals may be more efficient.

From a theoretical viewpoint, it is important to understand the mathematical properties of the stochastic processes introduced here, the adaptive measurement and the measurement trajectory. Finally, it is intriguing to consider to what extent the proposed method can adapted to multi-parameter estimation and to general (time-dependent) matrix product states, as opposed to stationary output states of Markov processes.

\vspace{4mm}

\noindent
{\bf Acknowledgements:} MG was supported by the EPSRC grant EP/T022140/1.

%
%
\section{Appendix: Derivation of the measurement filter}
\label{app:filter.derivation}

We start by applying the algorithm \citep{Zhou20} to our estimation problem. The natural setup is to measure the noise units in the order in which they emerge in the output, followed by a final measurement on the system+absorber. For simplicity we refer to the latter as the system. We start by defining
\begin{equation}\label{eq:Mn}
    M^{(n)} \coloneqq \ket{\Psi(n)}\bra{\dot{\Psi}(n)} - \ket{\dot{\Psi}(n)}\bra{\Psi(n)},
\end{equation}
where the index $n$ keeps track of the output length. In this section we will use the label $0$ for  the system, and $1,\dots ,n$ for the noise units of the output.
The algorithm prescribes measurement bases $\left\{ |e_i^{[l]} \rangle\right\}_{i=1}^k$ for each of the output units $l=1,\dots , n$ in an adaptive, sequential fashion. The first basis satisfies the equations
$$
\left\langle e^{[1]}_{i} \right| M^{(n)}_{1} \left| e^{[1]}_{i}\right\rangle = 0
\qquad \text{and} \qquad 
\left| \left\langle \left. e^{[1]}_{i}  \right| \chi \right\rangle  \right|^2=\frac{1}{k},
\qquad {\rm for~all~} i.
$$
where $M^{(n)}_{1}= {\rm Tr}_{0,2,\dots ,n} M^{(n)}$. The second equality follows from the fact that at $\theta_0$ we have $|\Psi(n)\rangle = |\psi\otimes \chi^{\otimes n}\rangle$. The next basis depends on the outcome $i_1$ of the first measurement and satisfies the constraints 
$$
\left\langle e^{[2]}_{i} \right| M^{(n)}_{2}(i_1) \left| e^{[2]}_{i}\right\rangle = 0
\qquad \text{and} \qquad 
\left| \left\langle \left. e^{[2]}_{i}  \right| \chi \right\rangle  \right|^2=\frac{1}{k},
\qquad {\rm for~all~} i
$$
where 
$$
M^{(n)}_{2}(i_1) = {\rm Tr}_{0,3,\dots, n} \left\langle e^{[1]}_{i_1} \right| M^{(n)} \left| e^{[1]}_{i_1}\right\rangle .
$$

Assuming the first $j<n$ units have been measured and a measurement record $i_{\underline{j}}:=\{ i_1, \dots , i_j\}$ has been obtained, we denote
$$
    M^{(n)}(\underi{j}) \coloneqq  
    \left
    \langle 
    e^{[1]}_{i_1} \otimes e^{[2]}_{i_2} \otimes \dots \otimes e^{[j]}_{i_j} \right| M^{(n)} \left|e^{[1]}_{i_1} \otimes e^{[2]}_{i_2} \otimes \dots \otimes e^{[j]}_{i_j}\right\rangle,
    \qquad M^{(n)}_{j+1}(\underi{j}) := {\rm Tr}_{0,j+2, \dots, n} M^{(n)}(\underi{j}).
$$
The measurement basis $\left\{\left| e^{[j+1]}_{i}(\underi{j})\right\rangle\right\} $ for unit $j+1$ is then obtained by solving the constraints
$$
\left\langle e^{[j+1]}_{i}(\underi{j}) \right| M^{(n)}_{j+1}(\underi{j}) \left| e^{[j+1]}_{i}(\underi{j}) \right\rangle = 0
\qquad \text{and} \qquad 
\left| \left\langle \left. e^{[j+1]}_{i}(\underi{j})  \right| \chi \right\rangle  \right|^2
=\frac{1}{k},
\qquad {\rm for~all~} i.
$$
The last step consists of measuring the system using the same procedure as for the output units.


A priori, the procedure depends on the size $n$ of the output. The following lemma shows that the optimal bases obtained for different output sizes coincide.

\begin{lemma}
Let $j,n$ be two output lengths with $j<n$ and consider applying the above procedure to the corresponding states $|\Psi^{(j)}\rangle$ and respectively $|\Psi^{(n)}\rangle$. The optimal measurement bases 
$\{ | e^{[l]}_i \rangle \}$ for the units $l= 1,\dots , j$ satisfy the same constraints and can be chosen to be the same.
In addition we have
$$
M^{(n)}_j(\underi{j-1}) = M^{(j)}_j(\underi{j-1}).
$$

\end{lemma}

\par

\noindent
\emph{Proof.} Recall that 
$|\Psi(n)\rangle = W^{(n)}\dots W^{(1)} |\psi\otimes \chi^{\otimes n}\rangle$, and let us denote 
$$
P(n):=\ketbra{\psi\otimes\chi^{\otimes n}} = |\Psi_{\theta_0}(n) \rangle \langle \Psi_{\theta_0}(n)|.
$$ 
Then 
$$
|\dot{\Psi}(n)\rangle = \sum_{i=1}^n
W^{(n)}\dots \dot{W}^{(i)}\dots W^{(1)}|\psi\otimes \chi^{\otimes n}\rangle= 
\sum_{i=1}^n
W^{(n)}\dots \dot{W}^{(i)}|\psi\otimes \chi^{\otimes n}\rangle
$$
where we used the fact that 
$W$ leaves $|\psi\otimes \chi\rangle$ invariant.

We first show that the matrix $M_1^{(n)}$ does not depend on $n$, and therefore the first measurement basis does not depend on the length of the output. 

\begin{align}
   M^{(n)}_1= {\rm Tr}_{0,2,\dots,n}\left[ M^{(n)} \right] &= \sum_i \text{Tr}_{0,2,\dots,n} \left[ \ W^{(n)}\cdots W^{(i)} \cdots W^{(1)} P(n) W^{(1)*} \cdots\dot{W}^{(i)*} \cdots W^{(n)*} - c.c \ \right] \nonumber\\
    &= \sum_i \text{Tr}_{0,2,\dots,n} \left[ \ W^{(n)} \cdots W^{(i)} P(n) \dot{W}^{(i)*} \cdots W^{(n)*} - c.c \ \right] \nonumber \\
    &= \sum_i \text{Tr}_{0,2,\dots,n} \left[ \ W^{(i)} P(n) \dot{W}^{(i)*} - c.c \ \right] \nonumber \\
    &= {\rm Tr}_{0} \left[ W^{(1)} P(1) \dot{W}^{(1)*} - c.c \right] + \sum_{i=2} \text{Tr}_{0,2,\dots,n} \left[ W^{(i)} P(n) \dot{W}^{(i)*} - c.c \right] \nonumber\\
    %
    %
    &= {\rm Tr}_{0} \left[M^{(1)}\right] =M^{(1)}_1, \label{eq:level1}
\end{align}
where $c.c,$ denotes the adjoint. 
In the third equality we used 
$$
{\rm Tr}_{0,2}[W_{0,2} A_{0,1,2} W^*_{0,2}]= {\rm Tr}_{0,2}[A_{0,1,2}]
$$
where $W_{0,2}$ is a unitary acting on subsystems $0,2$ of a tripartite system, and $A_{0,1}$ acts on subsystems $0,1$. In the last equality we used our assumption $\bra{\psi\otimes\chi}W^*\dot{W}\ket{\psi\otimes\chi}=0$. 
\par
Let $\left\{ \left| e^{[1]}_i \right\rangle \right\}$ be the measurement basis determined from $M_1= M_1^{(n)}= M_1^{(j)}$. We now show that, conditional on the outcome $i_1$ of this measurement, the second basis $\left\{ \left|e^{[2]}_i\right\rangle \right\}$ does not depend on the length of the output, which can be take to be equal to 2.
\begin{align*}
M_2^{(n)}(i_1)&=
    {\rm Tr}_{0,3,\dots ,n} \left[M^{(n)}(i_1)\right] \\
    &= \sum_i {\rm Tr}_{0,3,\dots,n} \left[ \left\langle e^{[1]}_{i_1} \right| W^{(n)}\cdots W^{(i)}\cdots W^{(1)} P(n) W^{(1)*}\cdots \dot{W}^{(i)*}\cdots W^{(n)*}  \left|e_{i_1}^{[1]}\right\rangle - c.c \right] \\
    &= {\rm Tr}_{0,3,\dots ,n} \left[ \left\langle e^{[1]}_{i_1} \right| W^{(n)}\cdots W^{(1)} P(n) \dot{W}^{(1)*} \cdots W^{(n)*} \left|e_{i_1}^{[1]}\right\rangle - c.c  \right] \\
    & \quad 
    +  {\rm Tr}_{0,3,\dots ,n}  \left[ \left\langle e^{[1]}_{i_1} \right| W^{(n)}\cdots W^{(2)}W^{(1)} P(n) W^{(1)*}\dot{W}^{(2)*}\cdots W^{(n)*}  \left|e_{i_1}^{[1]}\right\rangle - c.c \right] \\
    & \qquad
    +  \sum_{i=3}^n {\rm Tr}_{0,3,\dots ,n} \left[ \left\langle e^{[1]}_{i_1} \right| W^{(n)}\cdots W^{(i)}\cdots W^{(1)} P(n) W^{(1)*}\cdots \dot{W}^{(i)*}\cdots W^{(n)*}  \left|e_{i_1}^{[1]}\right\rangle- c.c \right] \\
    &= {\rm Tr}_{0} \left[ W^{(2)} \left\langle e^{[1]}_{i_1} \right| W^{(1)} P(2) \dot{W}^{(1)*} \left|e_{i_1}^{[1]}\right\rangle W^{(2)*} - c.c\right] \\
    & \quad
    +  \frac{1}{k} {\rm Tr}_{0} \left[ W^{(1)} P(1) \dot{W}^{(1)*} - c.c \right] \\
    &= {\rm Tr}_{0} \left[ W^{(2)} \left(\left\langle e^{[1]}_{i_1} \right| M^{(1)} \left|e_{i_1}^{[1]}\right\rangle \otimes 
    P_\chi \right) W^{(2)*} \right] + \frac{1}{k}{\rm Tr}_{0} \left[ M^{(1)} \right] =M_2^{(2)}(i_1).
\end{align*}
The equalities follow in the same way as in \eqref{eq:level1}, and in addition we used $\left|\braket{e^{[1]}_{i_1} | \chi}\right|^2 = \frac{1}{k}$ in the third equality. 

Using the same techniques as above we obtain the general statement
\begin{align}
  M_j^{(n)}(i_{\underline{j-1}})=  {\rm Tr}_{0,j+1,\dots ,n}\left[ M^{(n)}(\underi{j-1}) \right] &=  
    \frac{1}{k^{j-1}} {\rm Tr}_{0} \left[ M^{(1)} \right] \nonumber\\
    &+ {\rm Tr}_{0} \left[ W^{(j)} \left(\bra{e^{[\underline{j-1}]}_{\underi{j-1}}} M^{(j-1)}(\underi{j-2}) \ket{e^{[\underline{j-1}]}_{\underi{j-1}}} \otimes P_\chi \right) W^{(j)*} \right] \nonumber\\
    &= {\rm Tr}_{0} \left[ M^{(j)}(\underi{j-1}) \right] = M_j^{(j)}(i_{\underline{j-1}}).
    \label{eq:mjn.mjj}
\end{align}

\qed


Next, we show that we can express $M^{(j)}(\underi{j-1})$ in terms of $M^{(1)}$, $M^{(j-1)} (\underi{j-2})$ and $\ket{ e^{[j-1]}_{ j-1 } }$. Indeed by writing 
$W(j) = W^{(j)} W(j-1)$ we have
\begin{align}
    M^{(j)}(\underi{j-1}) &:= \left\langle e^{[\underline{j-1}]}_{ \underi{j-1} } \right| 
    M^{(j)} 
    \left| e^{[\underline{j-1}]}_{ \underi{j-1} } \right\rangle \nonumber \\
   &= \left\langle e^{[\underline{j-1}]}_{ \underi{j-1} } \right| 
    W^{(j)}W(j-1) P(j) W^*(j-1)\dot{W}^{(j)*} 
    \left| e^{[\underline{j-1}]}_{ \underi{j-1} } \right\rangle  -c.c
    \nonumber\\
    & \qquad + \left\langle e^{[\underline{j-1}]}_{ \underi{j-1} } \right|
    W^{(j)}W(j-1) P(j)\dot{W}(j-1)^*W^{(j)*} 
    \left| e^{[\underline{j-1}]}_{ \underi{j-1} } \right\rangle  -c.c
    \nonumber\\
    & = \frac{1}{k^{j-1}} M^{(1)} +
    W^{(j)}\left(\left\langle e^{[j-1]}_{i_{j-1}} \right|
M^{(j-1)}(i_{\underline{j-2}})\left| e^{[j-1]}_{i_{j-1}} \right\rangle\otimes P_\chi\right) W^{(j)*} \label{eq:mj}
\end{align}
This can then be used to determine the next measurement basis, producing the iterative procedure described in section \ref{sec:adaptive.measurement}, which consists in updating 
the `filter' that determines the optimal basis at each time step using the last measurement outcome $|e^{[j]}_{i_{j}}\rangle$. 

Let $A_1= M^{(1)}$ and  $A_j:= M^{(j)}(i_{\underline{j-1}})$ for $j>1$, and denote
$$
\Pi_j:= 
\left\langle 
e^{[j]}_{i_{j}}\right|
A_j\left| e^{[j]}_{i_{j}} \right\rangle=
\left\langle e^{[j]}_{i_{j}}\right|
M^{(j)}(\underi{j-1})\left| e^{[j]}_{i_{j}} \right\rangle
$$ 
Then equation \eqref{eq:mj} can be written as
$$
A_j = \frac{1}{k^{j-1}} A_1 + 
U^{(j)}\left(
\Pi_{j-1}\otimes P_\chi\right) U^{(j)*}
$$
The optimal measurement is obtained by applying the conditions to the operator $B_j := {\rm Tr}_0 A_j$.

\section{Appendix: Proof of Proposition \ref{prop:CFI-QFI}}

We start by using a general Fisher information identity for bipartite systems. Consider a generic pure state model $|\psi_\theta\rangle \in \mathcal{H}_s \otimes \mathcal{H}_o$ and let $\{|e^o_i\rangle\}$ and $\{|e^s_j\rangle\}$ be optimal bases in the `output' and `system' subsystems, for estimating $\theta$ at a particular value $\theta_0$, as prescribed by \citep{Zhou20}. Assume we perform the `output' measurement and let $X$ denote the outcome whose distribution is
$$
\mathbb{P}_\theta(X=i) =
\langle\psi_\theta | \mathbf{1} \otimes P_i| \psi_\theta\rangle, \qquad P_i = 
|e^o_i\rangle \langle e^o_i|.
$$
The conditional state of the `system' given $X=i$ is
$$
|\psi_\theta(i)\rangle = \frac{\mathbf{1} \otimes P_i |\psi_\theta\rangle }{\|\mathbf{1} \otimes P_i \psi_\theta \|}
$$
and this state contains the `remaining' information about $\theta$. The following inequality bounds the total available information as
$$
I_\theta(X) + \mathbb{E}_X F(\psi_\theta(X)) \leq F_\theta
$$
where the first term on the left side is the classical Fisher information of the outcome distribution 
$ \mathbb{P}_\theta$ and the second is the expected QFI of the conditional state 
$|\phi^i_\theta\rangle$. Consider 
now the second measurement $|e^s_j\rangle$ on the system and let $Y$ be its outcome. Then
$$
F_\theta = I_\theta(X) + 
\mathbb{E}_X I_\theta (Y|X) \leq I_\theta(X) + \mathbb{E}_X F(\psi_\theta(X)) \leq F_\theta
$$
where the first equality is due to measurement optimality, while the second is the inequality between classical and quantum information. This implies that 
\begin{equation}
\label{eq:bound.system.info}
F_\theta - I_\theta(X) = \mathbb{E}_X F_\theta (\psi_\theta(X)).
\end{equation}

We now consider the Markov setup in which the system+absorber play the role of `system' while the $n$ noise units are the `output'. We assume that the the output and the system+absorber are measured according to the optimal scheme presented in section \ref{sec:adaptive.measurement}. The joint state is given by 
$$
|\Psi_\theta (n)\rangle = W_{\theta}^{(n)}\cdot \dots \cdot 
W_{\theta}^{(1)} |\psi\otimes \chi^{\otimes n}\rangle 
$$
and the conditional states are 
$$
|\psi_\theta(i_1,\dots, i_n)\rangle= 
\frac{K^{[n]}_{\theta,i_n}\dots K^{[1]}_{\theta, i_1}
|\psi\rangle}
{\sqrt{p_\theta(i_1,\dots i_n)}}.
$$
We will show that at $\theta=\theta_0$ the left side of \eqref{eq:bound.system.info} is bounded by a constant which does not depend on $n$. For simplicity, whenever possible we will use the compact notations such as 
$$
K_{\bf i} := K^{[n]}_{\theta,i_n}\dots K^{[1]}_{\theta, i_1},\qquad 
\mathrm{and}\qquad 
|e_{\bf i}\rangle = |e^{[n]}_{i_n}\otimes\dots \otimes e^{[1]}_{i_1}\rangle.
$$
Recall that by design the following condition holds $W_{\theta_0}|\psi\otimes \chi\rangle = |\psi\otimes \chi\rangle$, which implies 
$
|\Psi_{\theta_0} (n)\rangle 
= 
|\psi\otimes 
\chi^{\otimes n}\rangle
$ 
and also 
$K^{[j]}_{i_j} |\psi\rangle  = c^{[j]}_{i_j} |\psi\rangle$ for some constants $c^{[j]}_{i_j}$.


Recall that for a pure state model $|\psi_\theta\rangle$ the QFI is given by 
$$
F_\theta= 4\left(\|\dot{\psi}_\theta\|^2 - 
|\langle \dot{\psi}_\theta | \psi_\theta\rangle |^2\right) = 4 \|\psi_\theta^\perp\|^2, \qquad |\psi_\theta^\perp \rangle= |\dot{\psi}_\theta\rangle - P_{\psi_\theta} |\dot{\psi}_\theta\rangle.
$$
Therefore, the expected system QFI on the left side of \eqref{eq:bound.system.info} is given by
\begin{equation}
\label{eq:QFI.system}
F_s(\theta_0) = 
4 \sum_{\bf i} p({\bf i}) 
\| \psi^\perp_{\theta_0} ({\bf i})\|^2
\end{equation}
where 
$$
|\psi^\perp_{\theta_0} ({\bf i})\rangle = 
|\dot{\psi}_{\theta_0} ({\bf i})\rangle - P_\psi| \dot{\psi}_{\theta_0} ({\bf i})\rangle
$$ 
For simplicity we now drop the subscript $\theta_0$ and we have
$$
|\dot{\psi} ({\bf i})\rangle = 
\frac{\dot{K}_{\bf i} |\psi\rangle}{\sqrt{p({\bf i})}} -\frac{1}{2}\frac{K_{\bf i} |\psi\rangle}{p^{3/2}({\bf i})} \dot{p}({\bf i})=
\frac{\dot{K}_{\bf i} |\psi\rangle}{\sqrt{p({\bf i})}} -\frac{1}{2}\frac{c({\bf i})\dot{p}({\bf i}) }{p^{3/2}({\bf i})} |\psi\rangle
$$
and therefore
$$
|\psi^\perp ({\bf i})\rangle = 
\frac{1}{\sqrt{p({\bf i})}} 
\left(I- P_\psi\right)
\dot{K}_{\bf i} |\psi\rangle. 
$$
Equation \eqref{eq:QFI.system} becomes
$$
F_s(\theta_0) = 4\sum_{\bf i} 
\left\| P^\perp_\psi \dot{K}_{\bf i} \psi\right\|^2= 
4 \left\| (P^\perp_\psi\otimes I) \dot{\Psi}(n)\right\|^2 .
$$
Now 
$$
|\dot{\Psi}(n)\rangle= 
\sum_{j=1}^n
W^{(n)}\dots W^{(j+1)}\dot{W}^{(j)} W^{(j-1)}\dots W^{(1)}|\psi\otimes \chi^{\otimes n}\rangle=
\sum_{j=1}^n W^{(n)}\dots W^{(j+1)}\dot{W}^{(j)}|\psi\otimes \chi^{\otimes n}\rangle
$$
and by triangle inequality
$$
F_s(\theta_0)\leq 4
\left(\sum_{j=1}^n \left\|(P^\perp_\psi\otimes I) W^{(n)}\dots W^{(j+1)}\dot{W}^{(j)}\psi\otimes \chi^{\otimes n} \right\| \right)^2
$$
Let $|\dot{\Psi}(1)\rangle := \dot{W}|\psi\otimes \chi\rangle$ 
and let $\tau:= {\rm Tr}_1(\dot{\Psi}(1)\rangle\langle \dot{\Psi}(1)|) $. Then
\begin{eqnarray*}
\left\|(P^\perp_\psi\otimes I) W^{(n)}\dots W^{(j+1)}\dot{W}^{(j)}\psi\otimes \chi^{\otimes n} \right\|^2 &=&
{\rm Tr}_{0,1,\dots n-j}\left((P^\perp_\psi\otimes I) W^{(n-j)}\dots W^{(1)} \tau W^{(1)*} \dots W^{(n-j)*} \right) \\
&=& {\rm Tr}_0(P^\perp_\psi T^{n-j} (\tau))
\end{eqnarray*}
where in the last equality we have uses the definition of the transition operator $T$ of the system+absorber. Assuming that $T$ is ergodic we have
$$
T^{n} (\tau) \to P_\psi
$$
exponentially fast with $n$ so that 
$$
{\rm Tr}_0(P^\perp_\psi T^{n-j} (\tau))\leq a^{2(n-j)}
$$
for some $a<1$. Therefore 
$$
F_s(\theta_0)\leq 4 
\left(\sum_{j=1}^n a^{n-j}\right)^2\leq 4\frac{1}{(1-a)^2}
$$
Note that if the spectral gap of $T$ becomes small then the convergence to stationarity is slower and the upper bound increases.

\qed

\section{Appendix: Proof of Lemma \ref{lemma:CFI-simulations}}
\label{app:proof_lemma_simulations}

The CFI of any output measurement can be computed explicitly by writing
\begin{eqnarray*}
\left.\frac{d}{d\theta}
p_\theta(i_1,\dots i_n) \right|_{\theta_0} &=&
\frac{d}{d\theta}\left. \left\|K^{[n]}_{i_n} \cdots K^{[1]}_{i_1}\psi\right\|^2 \right|_{\theta_0} \\
&=&
2{\rm Re}
\sum_{j=1}^n 
\langle \psi| K^{[1]*}_{i_1} \cdots 
K^{[n]*}_{i_n}
K^{[n]}_{i_n}\cdots \dot{K}^{[j]}_{i_j}\cdots K^{[1]}_{i_1}|\psi\rangle\\
&=&
\left|c^{[1]}_{i_1}\cdots c^{[n]}_{i_n}\right|^2\cdot  2{\rm Re}\sum_{j=1}^n 
\frac{\langle \psi|K^{[n]}_{i_n}\dots K^{[j+1]}_{i_{j+1}} \dot{K}^{[j]}_{i_j}|\psi\rangle }{c^{[j]}_{i_{j}} \dots c^{[n]}_{i_n}}
\end{eqnarray*}

Therefore
\begin{eqnarray}
I^{({\rm out})}_{\theta_0}(n) &=&\sum_{i_1,\dots, i_n} \left|c^{[1]}_{i_1}\cdots c^{[n]}_{i_n}\right|^2 
\left( 2{\rm Re}\sum_{j=1}^n 
\frac{\langle \psi|K^{[n]}_{i_n}\dots K^{[j+1]}_{i_{j+1}} \dot{K}^{[j]}_{i_j}|\psi\rangle }{c^{[j]}_{i_{j}} \dots c^{[n]}_{i_n}}\right)^2 \nonumber\\
&=&
\sum_{i_1,\dots, i_n} 
p_{\theta_0}(i_1,\dots, i_n) f^2(i_1,\dots, i_n) = \mathbb{E}_{\theta_0} (f^2)
\end{eqnarray}
where $f$ is the function 
\begin{equation*}
f(i_1,\dots , i_n) = 2{\rm Re}\sum_{j=1}^n 
\frac{\langle \psi|K^{[n]}_{i_n}\dots K^{[j+1]}_{i_{j+1}} \dot{K}^{[j]}_{i_j}|\psi\rangle }{c^{[j]}_{i_{j}} \dots c^{[n]}_{i_n}}
\end{equation*}

\qed

\section{Appendix: Computation of finite time system-output QFI}
\label{app:QFI-finite-time}

From \eqref{eq.Psi_n} and \eqref{eq:QFI.pure} we have
\begin{eqnarray}
F^{({\rm s+o})}_\theta(n) &=& 4 \|\dot{\Psi}_\theta(n)\|^2
\nonumber \\
&=&4 \sum_{i_1,\dots i_n}
\left\| \sum_{j=1}^n  K_{i_n} \dots \dot{K}_{i_j} \dots K_{i_1} \psi\right\|^2 
\label{eq:QFI.sum}
\end{eqnarray}
where $K_i$ are the (fixed) Kraus operators with respect to the standard basis, and $|psi\rangle = |0\rangle$. Our specific model ha the feature that both $K_i$ and $\dot{K}_i$ map the basis vectors into each other:
\begin{align*}
K_0 |0\rangle  
&=|0\rangle 
&
\dot{K}_0 |0\rangle 
&= i|1\rangle  
&
K_0 |1\rangle  
&=\sqrt{1-\lambda}|1\rangle
&
\dot{K}_0 |1\rangle 
&= i\sqrt{1-\lambda}|0\rangle
\\ 
K_1 |0\rangle  
&= 0  
&
\dot{K}_1 |0\rangle 
&=|1\rangle 
&
K_1 |1\rangle  
&=\sqrt{\lambda} e^{i\phi}|0\rangle
&
\dot{K}_1 |1\rangle 
&= 0
\end{align*}
This allows to compute the QFI explicitly by noting the terms in the sum \eqref{eq:QFI.sum} with more that two indices equal to $1$ have zero contribution. The remaining terms can be computed as follows. The term with only zero indices is
\begin{equation}
\label{eq:QFI.term0}
F^{(0)} = 4\left\| \sum_{j=1}^n  K_{0} \dots \dot{K}_{0} \dots K_{0} |0\rangle \right\|^2 =
4\left|\sum_{j=1}^n i a^{n-j}\right|^2 = 
4\left(\frac{1-a^n}{1-a }\right)^2 
\end{equation}
where $a =\sqrt{1-\lambda}$.

Consider now a sequence $(0, \dots, ,0, 1,0, \dots 0 )$ with a single one on position $l$. Since 
$K_0 \dots \dot{K}_0\dots  K_1\dots K_0|0\rangle =0$ the only contributing terms will be those with derivative on the first $(l-1)$ $K_0$s or on $K_1$. This gives 
\begin{eqnarray}
F^{(1)} &=& 4\sum_{l=2}^n 
\left\| i \sum_{r=1}^{l-1} e^{i\phi} a^{l-1-r} b |0\rangle + a^{n-l}|1\rangle \right\|^2 + 
4\| a^{n-1} |1\rangle \|^2\nonumber \\
&=& 4  \sum_{l=2}^n 
\left(
b^2 \frac{1-a^{l-1}}{1-a} + a^{2(n-l)}  
\right) +  4 a^{2(n-1)}\nonumber\\
&=&
4(n-1)\frac{b^2}{(1-a)^2} +
4\frac{a^2 -a^{2n}}{(1- a)^2} -8 \frac{b^2 (a - a^{n})}{(1-a)^3} + 4 \frac{1-a^{2(n-1)}}{b^2} +4 a^{2(n-1)}
\label{eq:QFI.term1}
\end{eqnarray}
where $b= \sqrt{\lambda}$.

Finally, consider the sequences of the type $(0,\dots 0,  1, 0 ,\dots 0,1,0\dots 0 )$ with 1s on positions 
$1\leq i< k \leq n$. In this case the nonzero contributions come from terms where the derivative is on positions $j=i$. The Fisher contribution is
\begin{eqnarray}
F^{(2)} &=& 
4 \sum_{1\leq i< k \leq n} 
\| e^{i\phi}\sqrt{\lambda} (1-\lambda)^{(k-i-1)/2} |0\rangle \|^2\nonumber \\
&=& 
4\sum_{1\leq i< k \leq n} b^2 a^{2(k-i-1)} = 
4(n-1) - 4\frac{a^2 -a^{2n}}{b^2}.
\label{eq:QFI.term2}
\end{eqnarray}
Adding together the contributions \eqref{eq:QFI.term0}, \eqref{eq:QFI.term1} and \eqref{eq:QFI.term2} we obtain the total QFI 
\begin{eqnarray}
F_\theta^{({\rm s+o})}(n) &=& F^{(0)}+F^{(1)}+F^{(2)}\nonumber 
\\
&& 
 =\frac{8n}{1-a}\nonumber\\
&&+
4\left[ 
\left(\frac{1-a^n}{1-a }\right)^2 -\frac{b^2}{(1-a)^2} +
\frac{a^2 -a^{2n}}{(1- a)^2} 
-2 \frac{b^2 (a - a^{n})}{(1-a)^3} +  \frac{1-a^{2(n-1)}}{b^2} 
+ a^{2(n-1)} \right] \nonumber\\
&&-4\left[  1 + \frac{a^2 -a^{2n}}{b^2}
\right]
\end{eqnarray}
where the leading term is consistent with the QFI rate formula \eqref{eq:QFI.rate.model}.

\end{document}